\pdfoutput=1
\documentclass[AMA,STIX1COL]{WileyNJD-v2}
\usepackage{stix}
\usepackage{graphicx}
\usepackage{sidecap}
\usepackage{tabularx}
\usepackage{threeparttable}
\usepackage{epsfig}
\usepackage{comment}
\newcommand{\lc}{\lowercase}

\articletype{}

\begin{document}

\title{\large Ajalon: Simplifying the Authoring of Wearable Cognitive Assistants}

\author[1]{Truong An Pham}

\author[2]{Junjue Wang}

\author[1]{Yu Xiao}

\author[3]{Padmanabhan Pillai}

\author[2]{Roger Iyengar}

\author[2]{Roberta Klatzky}

\author[2]{Mahadev Satyanarayanan}

\authormark{PHAM \textsc{et al}}

\address[1]{\orgdiv{School of Electrical Engineering}, \orgname{Aalto University}, \orgaddress{\state{Espoo}, \country{Finland}}}

\address[2]{ \orgname{Carnegie Mellon University}, \orgaddress{\state{Pittsburgh,  Pennsylvania}, \country{USA}}}

\address[3]{\orgname{Intel Labs}, \orgaddress{\state{Pittsburgh, Pennsylvania}, \country{USA}}}

\corres{Yu Xiao, Department of Communications and Networking, School of Electrical Engineering, Aalto University, Konemiehentie 2, 02150 Espoo, Finland. \\
	Email: yu.xiao@aalto.fi\\\\
	Mahadev Satyanarayanan, Computer Science Department,Carnegie Mellon University, 5000 Forbes Avenue, Pittsburgh, PA 15213, USA. \\
	Email: satya@cs.cmu.edu}

\abstract[Summary]
{

  Wearable Cognitive Assistance (WCA) amplifies human cognition in
  real time through a wearable device and low-latency wireless access
  to edge computing infrastructure.  It is inspired by, and broadens,
  the metaphor of GPS navigation tools that provide real-time
  step-by-step guidance, with prompt error detection and correction.
  WCA applications are likely to be transformative in education,
  health care, industrial troubleshooting, manufacturing, and many
  other areas.  Today, WCA application development is difficult and
  slow, requiring skills in areas such as machine learning and
  computer vision that are not widespread among software developers.
  This paper describes {\em Ajalon,} an authoring toolchain for WCA
  applications that reduces the skill and effort needed at each step
  of the development pipeline.  Our evaluation shows that Ajalon
  significantly reduces the effort needed to create new WCA
  applications.

}


\maketitle

\footnotetext{\textbf{Abbreviations:} WCA, wearable cognitive assistance; }

\section{I\lc{ntroduction}}
\label{sec:intro}

\begin{figure*}
\begin{tabular}{|p{0.29in}|p{0.95in}|p{3.32in}|p{0.84in}|p{0.8in}|}
\hline
App Name & Example Input Video Frame & Description & Symbolic \phantom{000} Representation & Example Guidance \\
\hline
\phantom{000} \textbf{Pool}     & \raisebox{-0.9\totalheight}{\psfig{file=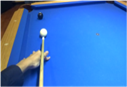, width=0.97in}}
&
Helps a novice pool player aim correctly. Gives continuous visual feedback (left arrow, right arrow, or thumbs up) as the user turns his cue stick.  The symbolic representation describes the positions of the balls, target pocket, and the top and bottom of cue stick.
&
\phantom{000} $<$Pocket, object ball, cue ball, cue top, cue bottom$>$ & \phantom{000} \raisebox{-0.85\totalheight}{\psfig{file=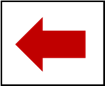, width=0.8in}} \\
\hline
\phantom{000} \textbf{Ping-pong} & \raisebox{-0.9\totalheight}{\psfig{file=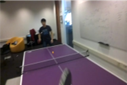, width=0.97in}}
&
Tells novice to hit ball to the left or right, depending on which is more likely to beat opponent. Uses color, line and optical-flow based motion detection to detect ball, table, and opponent.  Video URL: {\em \href{https://youtu.be/\_lp32sowyUA}{https://youtu.be/\_lp32sowyUA}}
&
\phantom{000} $<$InRally, ball position, opponent position$>$ & \phantom{000} Whispers ``Left!'' \\
\hline
\phantom{000} \textbf{Work-out} & \raisebox{-0.9\totalheight}{\psfig{file=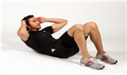, width=0.97in}}
&
Counts out repetitions in physical exercises. Classification is done
using Volumetric Template Matching on a 10-15 frame video segment.  A
poorly-performed repetition is classified as a distinct type of
exercise (e.g. ``good pushup'' versus ``bad pushup'').
&
\phantom{000} $<$Action, count$>$ & \phantom{000} Says ``8 '' \\
\hline
\phantom{000} \textbf{Face}     & \raisebox{-0.9\totalheight}{\psfig{file=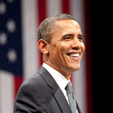, width=0.8in}}
&
Jogs your memory on a familiar face whose name you cannot recall. Detects and extracts a tightly-cropped image of each face, and then applies a state-of-art face recognizer. Whispers the name of the person recognized.
&
\phantom{000} ASCII text of name & \phantom{000} Whispers ``Barack Obama'' \\
\hline
\phantom{000} \textbf{Lego} & \raisebox{-0.9\totalheight}{\psfig{file=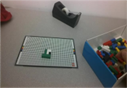, width=0.9in}}
&
Guides a user in assembling 2D Lego models.  The symbolic representation is a matrix representing color for each brick.   Video URL: {\em \href{https://youtu.be/7L9U-n29abg}{https://youtu.be/7L9U-n29abg}}
&
\phantom{000} [[0, 2, 1, 1], \break [0, 2, 1, 6], \break [2, 2, 2, 2]] \break & \raisebox{-0.85\totalheight}{\psfig{file=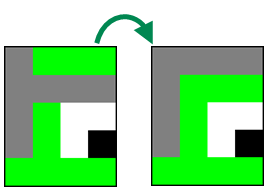, width=0.8in}} Says ``Put a 1x3 green piece on top'' \\
\hline
\phantom{000} \textbf{Draw} & \raisebox{-0.9\totalheight}{\psfig{file=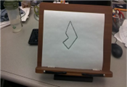, width=0.9in}}
&
Helps a user to sketch better. Builds on third-party app for desktops. Our implementation preserves the back-end logic. A new Glass-based front-end allows a user to use any drawing surface and instrument. Displays the error alignment in sketch on Glass.  Video URL: {\em \href{https://youtu.be/nuQpPtVJC6o}{https://youtu.be/nuQpPtVJC6o}}
&
\raisebox{-0.85\totalheight}{\psfig{file=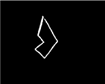, width=0.7in}} & \raisebox{-0.95\totalheight}{\psfig{file=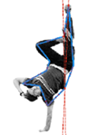, width=0.7in}} \\
\hline
\phantom{000} \textbf{Sand-wich} & \raisebox{-0.9\totalheight}{\psfig{file=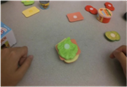, width=0.97in}}
&
Helps a cooking novice prepare sandwiches according to a recipe. Since real food is perishable, we use a food toy with plastic ingredients. Object detection uses faster-RCNN deep neural net approach. 

    Video URL: {\em \href{https://youtu.be/USakPP45WvM}{https://youtu.be/USakPP45WvM}}
&
\phantom{000} Object: \break ``E.g. Lettuce on top of ham and bread'' & \raisebox{-0.9\totalheight}{\psfig{file=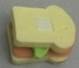, width=0.7in}} Says ``Put a piece of bread on the lettuce'' \\
\hline

\end{tabular}
\caption{Example Wearable Cognitive Assistance Applications (Source: Adapted from Satya~\cite{Satya2017})}
\label{fig:bigtable}
\vspace{-0.2in}
\end{figure*}

Since its introduction in 2014,\cite{ha2014towards} {\em Wearable
  Cognitive Assistance}~(WCA) has attracted considerable attention.  A
WCA application provides just-in-time guidance and error detection for
a user who is performing an unfamiliar task.  Even on familiar tasks,
prompt detection of errors can be valuable when the user is working
under conditions of fatigue, stress, or cognitive overload.
Informally, WCA is like having ``an angel on your
shoulder.''\cite{Ray2018}  It is inspired by, and broadens, the
metaphor of GPS navigation tools that provide real-time step-by-step
guidance, with prompt error detection and correction.
Figure~\ref{fig:bigtable} summarizes the attributes of a small sample
of the many WCA applications that we have built since 2014.  A YouTube video
of a WCA application to assemble an IKEA kit can be viewed at
\url{https://shorturl.at/iEOZ0}.

A WCA application runs on a wearable device such as Google Glass or
Microsoft Hololens, leaving the user's hands free for task
performance.  It provides visual and verbal guidance through the video
and audio channels of the wearable device.  It provides a user
experience that is similar to the ``look and feel'' of augmented
reality (AR) applications, while being deeply dependent on artificial
intelligence (AI) algorithms such as object recognition via computer
vision using deep neural networks (DNNs).  Because of the
computational limitations of lightweight wearable devices that have
acceptable battery life, a WCA application uses a wireless network to
offload its compute-intensive operations to a nearby {\em
  cloudlet}.\cite{Satya2009}  WCA applications are thus
simultaneously compute-intensive, bandwidth-hungry, and
latency-sensitive.  They are hence perceived as a class of ``killer
apps'' for {\em edge computing}.\cite{Satya2017,Satya2019c}

Based on our experience from authoring many WCA applications, this
paper describes a toolchain that we have built to simplify their
creation.  WCA applications are inherently task-specific because they
embody deep knowledge of the task being performed.  The task-specific
focus reduces the complexity of the WCA application, and prevents it
from becoming ``AI-Complete.''\cite{Yampoloskiy2011} There is a narrow
context within which sensor data is interpreted, progress to
completion is defined, and ambiguities are resolved.  For example, in
the task of assembling a kit of parts, the computer vision module only
needs to be able to accurately recognize the parts in the kit.
Everything else can be ignored as irrelevant information.  Because of
the narrowing of context, the WCA application is able to resolve
errors and provide detailed-enough instructions for a user to complete the task.

This paper focuses on a toolchain called {\em Ajalon} that simplifies
the task-specific aspects of authoring a WCA application.  Ajalon
layers task-specific components on top of a task-independent runtime
platform called {\em Gabriel} that we have released open
source.\cite{Gabriel2018,ha2014towards} For ease of exposition, we
will refer to WCA applications as ``Gabriel applications'' in the rest
of this paper.  Without Ajalon, authoring a Gabriel application is
time consuming, often taking multiple person-months of
effort.  This is clearly not scalable.  Ajalon's goal is to simplify
the process so that a small team (1-2 people) of a task expert and a
developer without computer vision expertise can create an initial
version of a Gabriel application within a few days.  This is a
productivity improvement of at least one order of magnitude.
Refinement and tuning may take longer, but can be guided by early use
of the application. We focus on vision-based Gabriel applications in
this paper, but plan to extend Ajalon to multi-sensor applications.

Ajalon has four stages:
\begin{itemize}
\item{ The first stage, {\em Preprocessing Tool (PT),} automatically
    extracts a workflow from videos that were created by task experts.
    Some of these videos may demonstrate common errors a novice
    typically makes.  PT automatically segments an input video into
    working steps.  At each step, a list of associated objects is
    discovered.  The segmentation into working steps and the list of
    associated objects are the output of PT.}

\item{In the second stage, the application developer refines the
    output of PT using {\em PTEditor.}  This refinement is needed because workflow
    extraction by PT is imperfect even with state-of-the-art computer
    vision techniques. PTEditor provides merge and split
    functions for the task expert to modify the steps of the workflow.
    The task expert can also edit the list of associated objects
    discovered by PT.}

\item{The third stage addresses the computer vision aspect of the
    toolchain.  It consists of associating task-specific {\em object
      detectors} based on DNNs for each object in the task.  In some
    cases, the necessary object detectors may already have been
    created and be available through a library.  This is likely to
    happen over time, as the use of WCA grows and object detectors
    become available for every manufactured component in a standard
    parts catalog.  For non-standard components and in the interim
    while WCA is still in its infancy, the creation of custom object
    detectors will be essential.  For this purpose, our toolchain
    includes {\em OpenTPOD.}  This tool enables even a developer who
    has no skill in computer vision or machine learning to easily
    create custom object detectors without writing a single line of
    code.  At the end of this Ajalon stage, object detectors are
    available for all task-relevant objects.}

\item{In the fourth and final stage of Ajalon, the developer uses a
    finite state machine (FSM) editor called {\em OpenWorkflow} to
    bring together all the pieces from the earlier stages.
    OpenWorkflow models user actions as transitions in a task-specific
    FSM.  Each state represents partial completion of the task, or an
    error.  Object detectors are used to detect the current state of
    the task, and the transitions are inferred from final and prior
    state.  To trigger the next desired transition, the developer can
    add visual and verbal guidance to a state.  At runtime, when the
    FSM enters that state, the associated visual and verbal guidance
    will be presented to the user.  The output of OpenWorkflow is
    executable code for the new Gabriel task.}
\end{itemize}

We describe Ajalon in detail in the rest of the paper.
Section~\ref{sec:background} provides background on Gabriel.  The four
stages of Ajalon are presented in depth in
Sections~\ref{sec:preprocessing} to \ref{sec:statemachine}.
Section~\ref{sec:evaluation} presents our evaluation of Ajalon through
microbenchmarks of individual components, as well as an end-to-end
user study of the entire toolchain.

\section{B\lc{ackground} \lc{and} R\lc{elated} W\lc{ork}}
\label{sec:background}
In this section, we introduce technical background and prior work on
the Gabriel platform, applications, and authoring process.
For brevity, we often shorten ``the Gabriel platform'' to just ``Gabriel''.

\begin{figure}
	\centering
	\includegraphics[width=0.89\linewidth, height=0.5\linewidth]{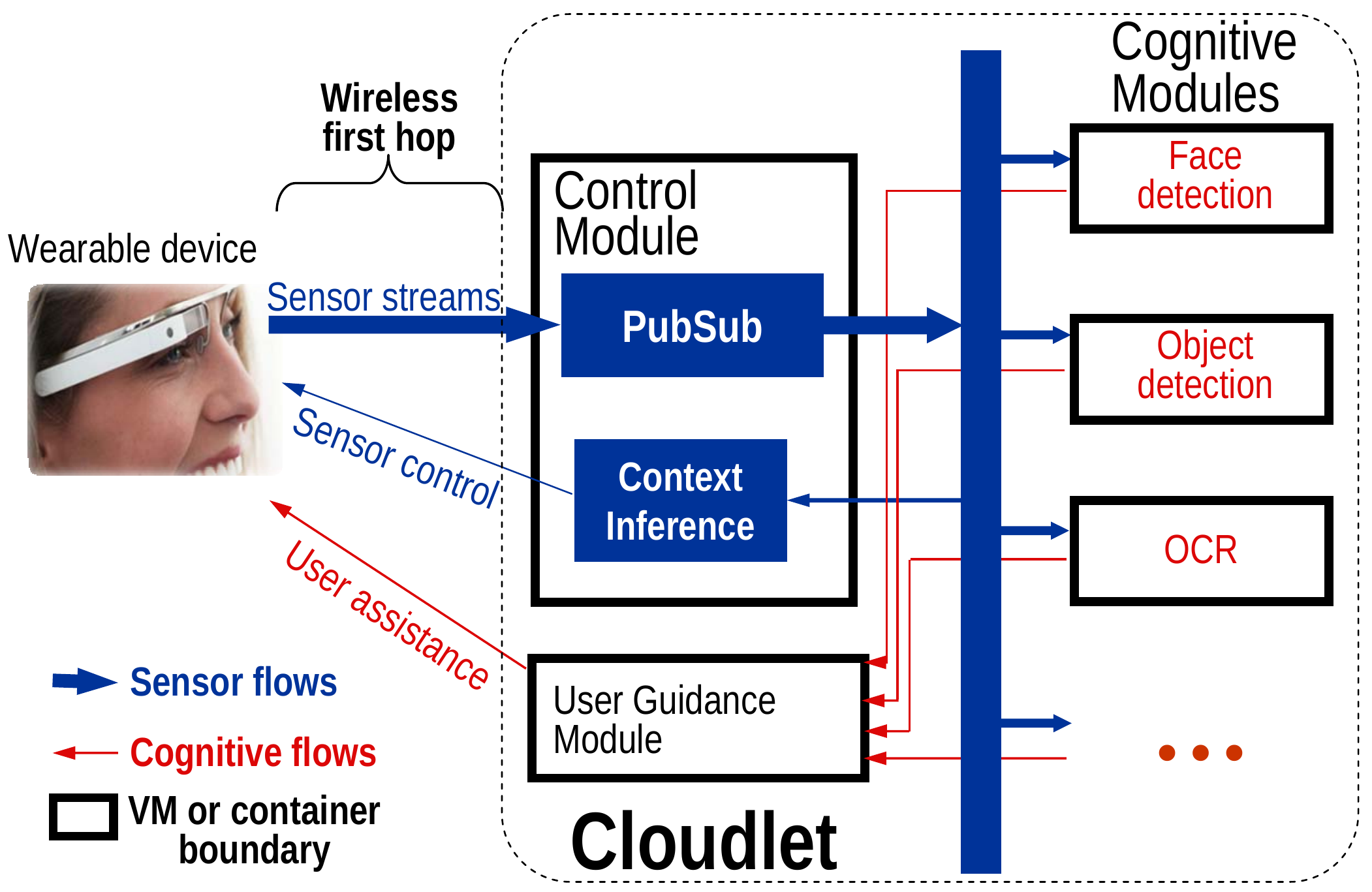}
	\caption{Gabriel Architecture {\rm\small{(Source: Satya et al.~\cite{Satya2019c})}}}
	\label{fig:gabriel}
	\vspace{-0.2in}
\end{figure}

\subsection{Gabriel Platform}
\label{sec:gabriel}

Gabriel enables its applications to preserve crisp quality of
experience (QoE) while overcoming the resource limitations of small,
lightweight, and energy-efficient mobile devices. It achieves this by
enabling a mobile device to offload compute-intensive operations to a
nearly {\em cloudlet} rather than to the distant
cloud.~\cite{Satya2009,Satya2017,Satya2019c} As shown in
Figure~\ref{fig:gabriel}, Gabriel is an extensible PaaS (Platform as a
Service) layer that we have created for WCA applications.  The
front-end on a wearable device performs preprocessing of sensor data
(e.g. compression and encoding), and then streams it over a wireless
network to a cloudlet.  The back-end on the cloudlet is organized as a
collection of {\em cognitive modules} that embody compute-intensive
algorithms such as object detection and speech recognition.  Depending
on the level of trust in a particular setting, each module can be
encapsulated within a virtual machine (VM), container, or process
boundary.  The {\em control module} in Gabriel is the focal point for
all interactions with the wearable device.  A publish-subscribe
(PubSub) mechanism decodes and distributes the incoming sensor
streams.  Cognitive module outputs are integrated by a task-specific
{\em User Guidance module} that performs higher-level cognitive
processing.  The application-specific code in the user guidance module
typically consists of a task state extractor and a guidance generator.
The task state extractor maps inputs from cognitive modules into the
user's current point of progress on the task.  Detection of this state
then triggers task-appropriate visual, verbal, or tactile guidance that
is transmitted back to the wearable device.  Since the Gabriel
back-end embodies all task-specific components, applications can be
easily ported to different wearable devices.  We have successfully
used a diversity of devices such as Google Glass, Microsoft HoloLens,
Vuzix Glass, and ODG R7.~\cite{chen2017}

\subsection{Authoring Gabriel Applications}
\label{sec:authoring}

\begin{figure}
	\centering
	\includegraphics[width=0.8\textwidth]{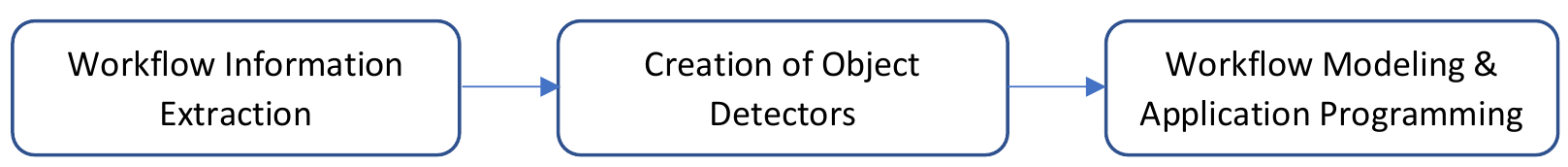}
	\caption{Current Gabriel applications development process}
	\label{fig:gabriel_current_dev_process}
	\vspace{-0.2in}
\end{figure}

Even using the Gabriel platform, WCA application development remains a
nontrivial endeavor.  The current development process involves three
main steps that are illustrated in
Figure~\ref{fig:gabriel_current_dev_process}: (a) workflow extraction,
(b) object detector creation, and (c) workflow modeling.  These steps
require very different types of expertise.  First, for workflow
extraction, a task expert is needed who understands how best to
perform the task at hand, as well as the potential mistakes someone
may make when trying to complete the task for the first time.  Second,
for creating task-specific object detectors using a deep neural
network (DNN), a computer vision expert is needed.  Object detection
is necessary for determining when specific steps of the task have been
completed. Note that the vision and task experts may need to iterate
on the task workflow to make it amenable to easily-trained and
accurate vision models.  Third, a software developer combines the
individual components into a front-end Android app that can be
published in the Google Play Store, and back-end software that can be
deployed on cloudlets.

Ajalon's goal is to simplify the process of creating a Gabriel
application so that a single developer, working closely with a task
expert, can create the application in a short time. This requires
creating higher-level abstractions, providing common reusable modules, and
reducing the amount of required computer vision and software
engineering knowledge.  This goal is different from recent efforts in
speeding up software development~\cite{rosenberg2019parallel} by
leveraging parallelism across several developers.  While such efforts
can shorten the elapsed time (i.e., wall clock time) for the
development process, it does not reduce the total number of
person-hours invested.  In contrast, Ajalon's goals are twofold. The first goal is reducing the total number of person-hours required for development. The second goal is enabling developers who do not have expertise in computer vision and machine learning.

\subsection{Workflow Extraction}
\label{sec:extraction}

Ajalon's goal of simplifying the creation of Gabriel applications has
some overlap with previous work on {\em workflow extraction.}  In this
context, the term ``workflow'' is defined as a machine-readable
description of a sequence of activities or working steps, that a user
must carry out in order to complete a specific task. Workflows, by
definition, are task-specific and often crafted in an ad-hoc fashion by task experts. Automatically extracting a workflow from a video depicting a task reduces application development time, especially for lengthy tasks with hundreds or even thousands of steps.

Previous research on workflow extraction was concerned with creating
tools for manual workflow extraction and computer vision algorithms
for automated video analysis.  The Computer Cooking Contest, which has
been run since 2008, challenges researchers to extract workflows from
cooking recipes available from sources such as Wikipedia. In an open challenge system called CookingCakeWf, researchers described
the steps required to make a recipe using a control flow, and they
described the ingredients and their products using a data
flow.~\cite{cooking2010a}  Mura et al.~\cite{mura2013ibes} proposed a tool
named IBES that facilitates manual working step separation and
annotation. Kim et al.~\cite{kim2014toolscape} developed
ToolScape, which helps verify workflow descriptions scripted by multiple people from videos depicting tasks. For automated video analysis, Nater et al.~\cite{nater2011u} developed an unsupervised
machine learning method to extract the workflow information from the
videos of an assembly line.  In contrast to these earlier efforts,
Ajalon leverages the fact that our headsets capture videos from a first-person perspective.

The current implementation of Ajalon assumes that the workflow of the
target application has specific simplifying attributes.  First, Ajalon
assumes a linear workflow that can be readily decomposed into a
sequence of steps. This is a common feature of tasks such as product
assembly or equipment servicing.  Second, Ajalon assumes that the task
follows a canonical workflow; i.e., there is a single preferred
procedure (or if several are comparable, one procedure can be
selected), and the steps in the task follow a prescribed order.
Third, Ajalon assumes that the correct completion of a step (and
transitively, all steps that precede it) can be visually determined
using object detection algorithms. This excludes, for example, a step
such as one that requires putting a small object inside of a large
object but leaves the large object visually unchanged.  In spite of
these simplifying assumptions, Ajalon is able to cover a wide range of
WCA applications.  Future versions of Ajalon may relax some of these
assumptions.

\section{A\lc{jalon} O\lc{verview}}
\label{sec:design}
\begin{figure*}[t]
	\centering
	\includegraphics[width=0.9\textwidth]{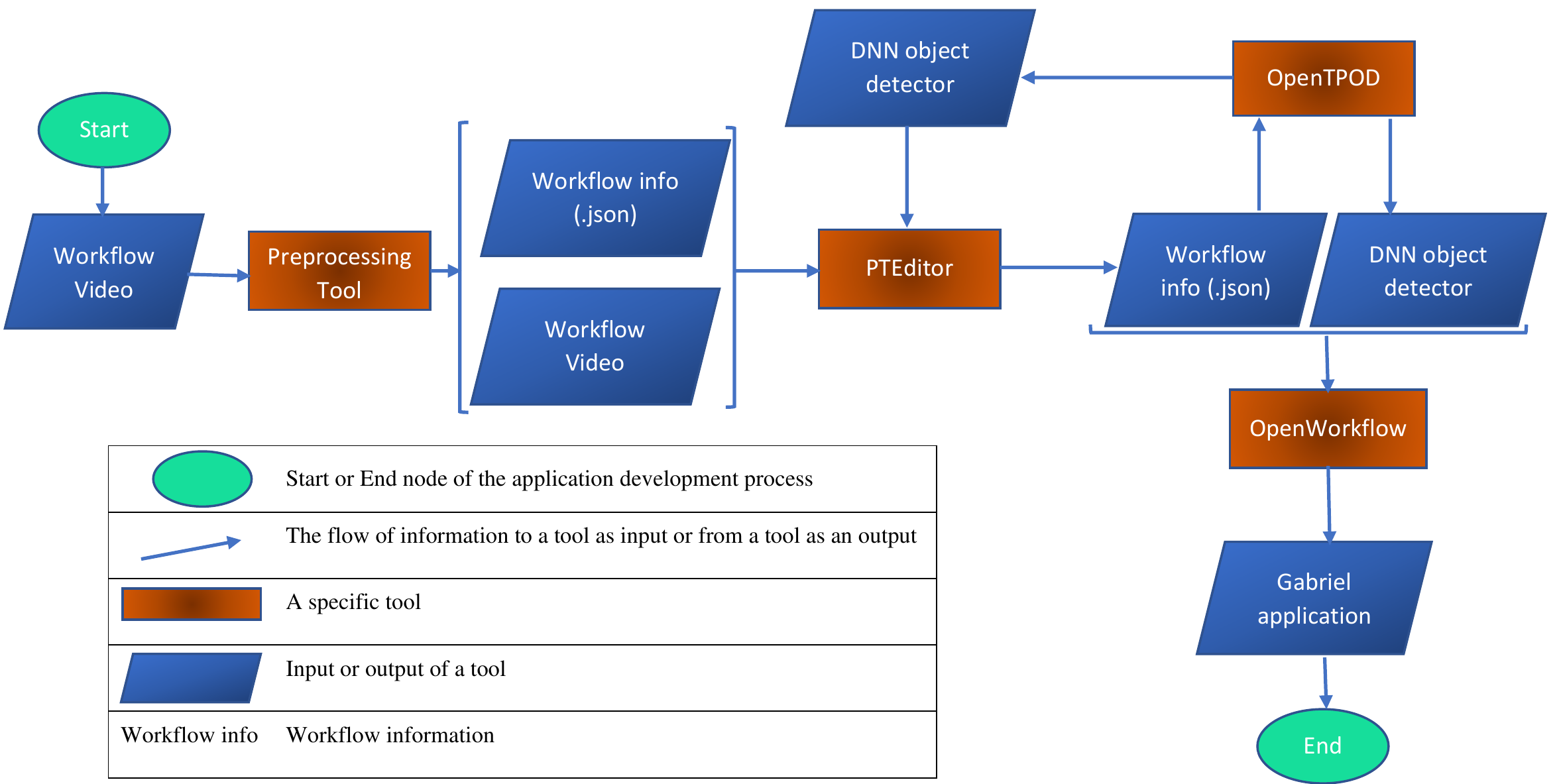}
	\caption{Ajalon pipeline overview}
	
	\label{fig:overview}
\end{figure*}

Ajalon enables a development team consisting of a task expert and a
software developer to jointly create a Gabriel application. We assume
that the task expert understands the steps of the task well, but has
no software skills.  In contrast, the developer is skilled in software
development, but is unfamiliar with the task.  Neither person needs to
be an expert in computer vision or machine learning, even though these
are central to Gabriel applications.  Ajalon codifies, simplifies and
partially automates the end-to-end authoring steps to help the
development team create a Gabriel application.

Figure~\ref{fig:overview} illustrates the four steps of the Ajalon
toolchain that were mentioned in Section~\ref{sec:intro}.  The process
starts by capturing an example video that shows a first-person
viewpoint of a task performed by an expert.  The video may also
illustrate common mistakes and guidance on how the expert fixes them.
Second, the development team uses Preprocessing Tool (PT) to automatically segment a video
into time-stamped working steps, and to generate a list of
task-relevant objects seen in the video.  The extracted set of working
steps is imported into PTEditor to manually correct mistakes made by
PT.  Third, the development team uses OpenTPOD to create a DNN-based
object detector that can accurately distinguish between all the
task-relevant objects discovered by PT.  Finally, the outputs of
PTEditor and OpenTPOD are used by OpenWorkflow to create an FSM whose
state transitions are triggered by visual changes in task state.  The output
of OpenWorkflow is executable code for the Gabriel application that
embodies the workflow of the task.

\begin{figure*}[t]
	\centering
	\includegraphics[width=0.95\textwidth]{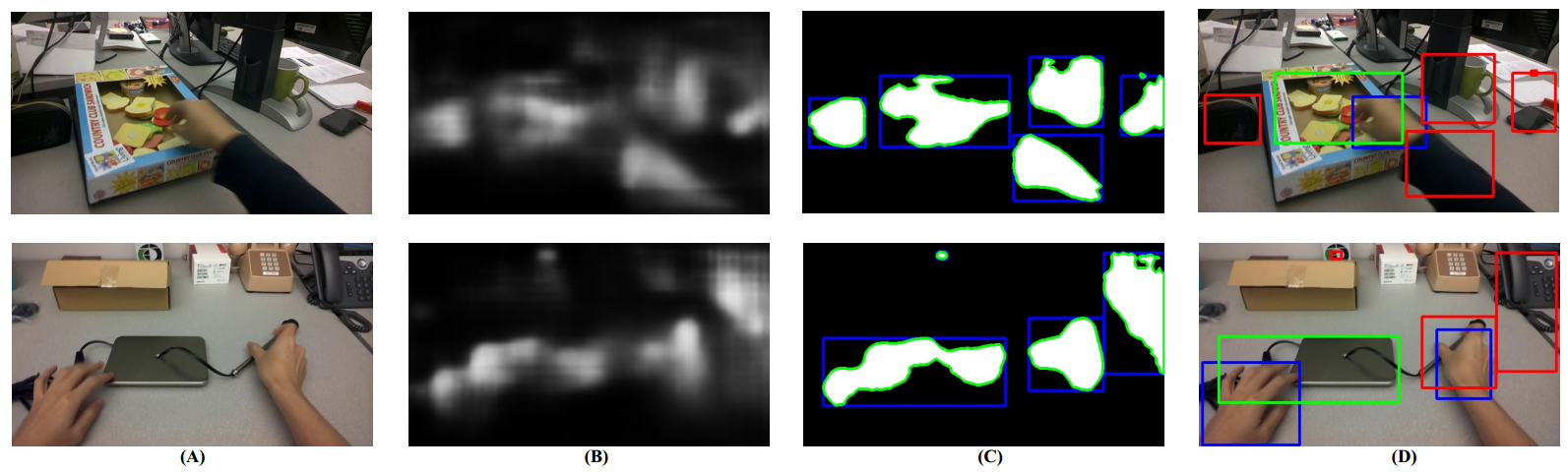}
	\caption{Working step separation.
          (A) Original frame.
          (B) EgoNet saliency detection.
          (C) RoIs bounding box extraction.
          (D) Hand-object interaction detection
        }
	\label{fig:workingstep_separation}
\end{figure*}

\section{PT: P\lc{reprocessing} T\lc{ool}}
\label{sec:preprocessing}

\begin{SCfigure}
	\centering
	\includegraphics[width=0.6\textwidth]{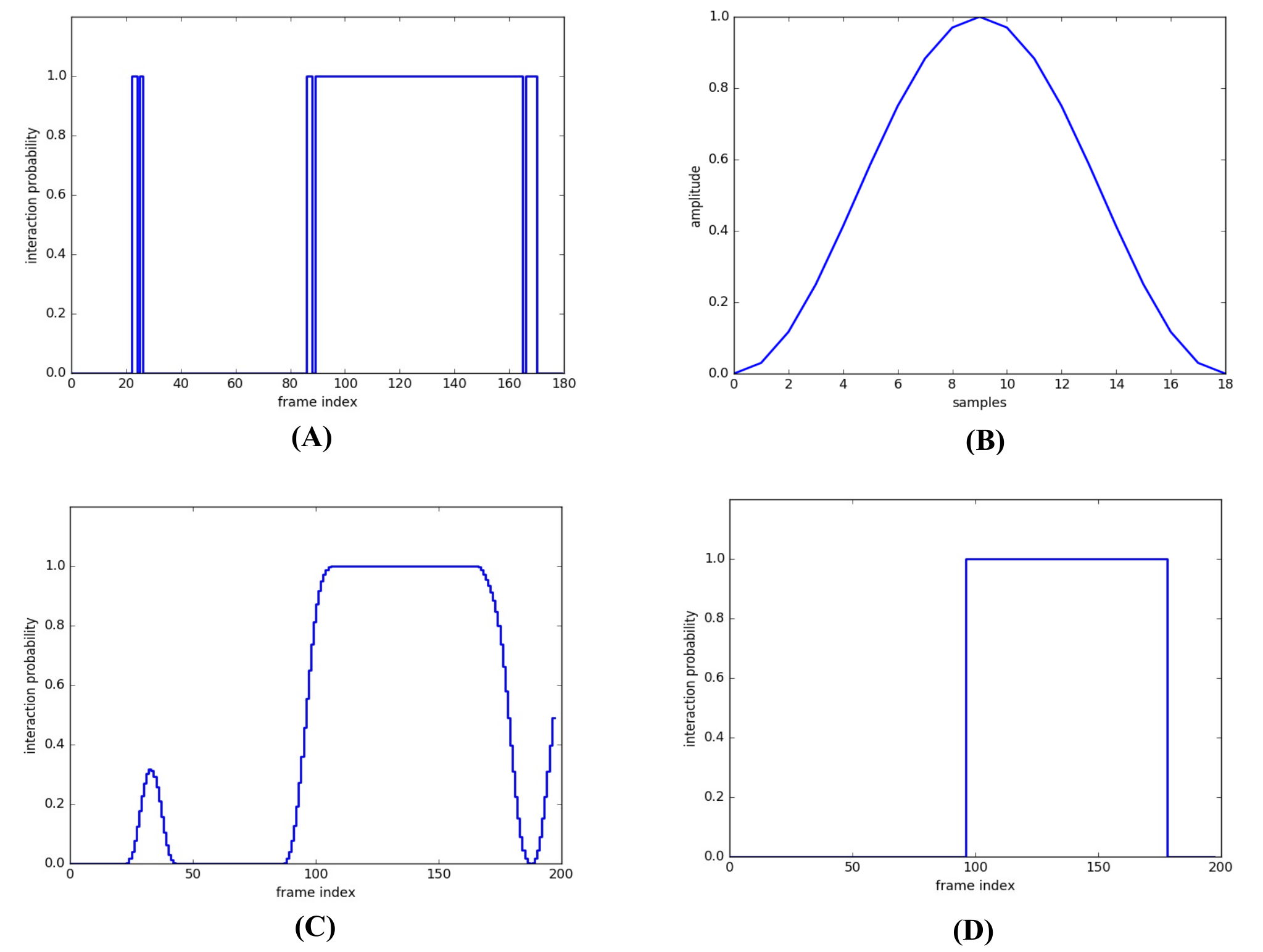}
	\caption{Noise reduction in working step detection.
          (A) Raw binary data.
          (B) Hanning window with size of 19.
          (C) Smoothing result.
          (D) Noise reduction after applying thresholding
        }
	\label{fig:interaction_smoothing}
\end{SCfigure}

\begin{SCfigure}
	\centering
	\includegraphics[width=0.6\textwidth]{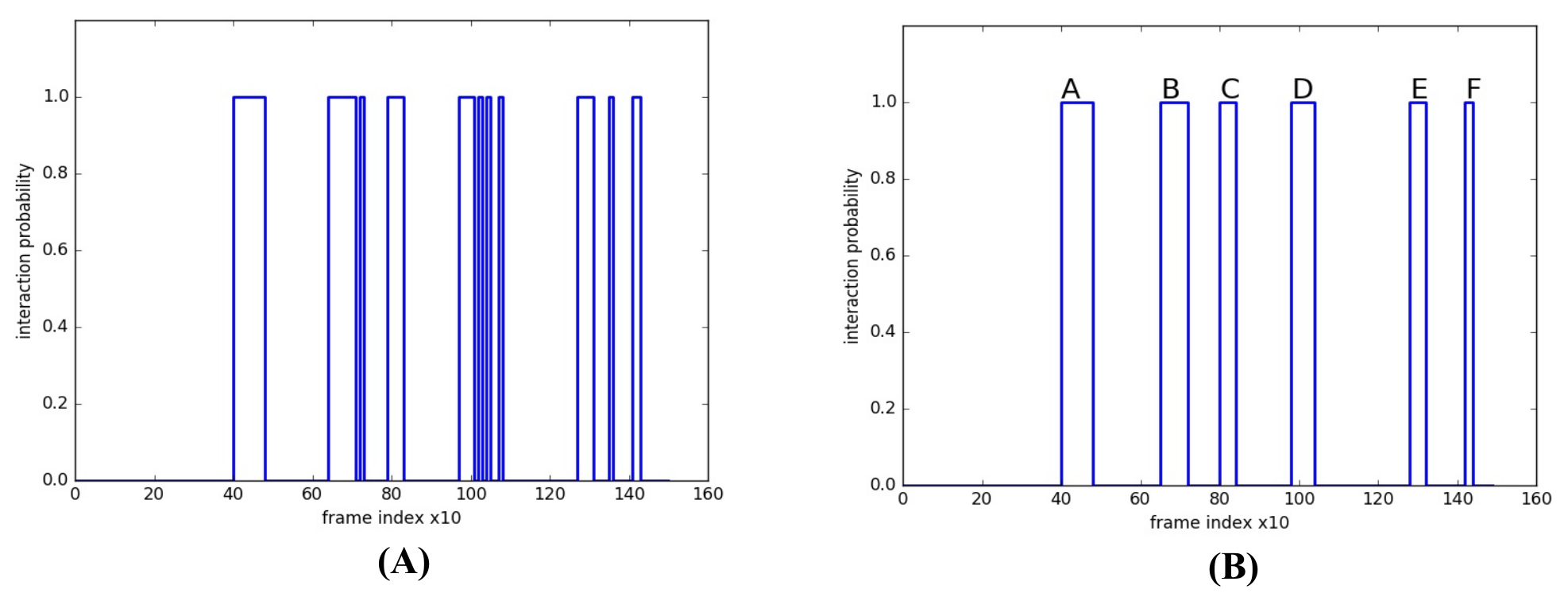}
	\caption{
          Refined working steps segmentation result on 1500 frames by smoothing
          technique.
          (A) Raw binary interaction result.
          (B) Refined sequence of working steps.
        }
	\label{fig:half_video_segmentation}
\end{SCfigure}

PT targets a manual assembly process in which parts are combined by
human hands. The input to PT is a first-person video of an assembly
process in which parts are added in sequence.  In such a video,
assembly actions are viewed as interactions between hands and objects
(i.e. parts or workpiece).  PT splits this video into segments, each
corresponding to a working step.  PT assumes that only one part is
added per step, and that each step changes the appearance of a
partially-assembled workpiece.  Step completion can therefore be
confirmed from workpiece appearance.  PT detects the boundary between
working steps by finding a lull in hand-object interactions.  We have
shown that this approach is reliable, and avoids the much more
difficult problem of reliably detecting specific hand
gestures.~\cite{pham2018unsupervised} The boundary detection process
works as follows:

\begin{enumerate}
\item Run EgoNet,~\cite{bertasius2016first} a two-stream network that
  predicts per-pixel likelihood of action-objects from first-person video,
  to obtain salient regions that represent the Regions of Interest (RoIs) of a
  video frame. Instead of going through each pixel within the salient regions,
  PT applies clustering and contour extraction to obtain the bounding
  boxes of RoIs, as illustrated in Figure~\ref{fig:workingstep_separation}.
  This narrows the search space for hand-object interaction to the few bounding boxes.

\item Run YOLOv3,~\cite{redmon2018yolov3} a real-time object detector,
  to detect human hands in each frame.  Assuming
  that the first-person video only captures the operations of one
  worker, the object detector is trained to detect the regions where
  the left and/or right hand of the worker appear.

\item Recognize hand-object interaction by detecting the overlap
  between hand regions and RoIs. After going through all frames, PT
  generates a one-dimensional binary array that represents the
  occurrences of interaction as a binary sequence. In
  Figure~\ref{fig:workingstep_separation}~D, the detected interaction
  regions are highlighted by green boxes, while hand regions and other
  RoIs are marked with blue and red boxes, respectively.

\item Smooth and threshold the interactions to reduce false positive
  and false negative errors in the 
  previous steps. For example, a false positive error may occur when
  an unintentional hand movement that has no interaction with objects is
  detected in an RoI. A false negative error may occur if PT misses a
  hand or RoI.  PT overcomes such errors by assuming continuity of
  hand-object interaction, and taking into account detection in
  neighboring frames. A smoothing technique is used to
  estimate the interaction between the interaction signal of a video
  and the frame index. As illustrated in Figure~
  \ref{fig:interaction_smoothing}, the interaction signal is
  represented as a probability of hand-object interaction along a
  video frame index. In Figure~\ref{fig:interaction_smoothing} A, an
  interaction detection result is visualized with only $1$ and $0$
  based on the detector above. 
  Then, based on our prior experience in extracting task steps from video, 
  we smooth the results by convolving with a Hanning window of size 19, 
  selected empirically as in other approaches to activity recognition.~\cite{jalal2017robust}  
  After smoothing the results, we use a threshold of 0.5
  to determine if an interaction happens at the considered frame. The convolution enhances the continuity of interaction detection and reduces noise. The smoothing result is
  illustrated in Figure~\ref{fig:interaction_smoothing}.

\item Cluster successive frames containing hand-object interactions to
  represent a continuous working step. The system eliminates small
  clusters that are  shorter than 12 frames, because we assume that
  the camera cannot capture a meaningful action in the corresponding
  time interval of less than half a second. The result of the entire
  process is illustrated in Figure~\ref{fig:half_video_segmentation}.
  Here, PT has segmented the video into 6 steps (labeled A-F) of intense
  hand-object interactions, separated by periods of no interactions 
  between hands and objects in the video.  
\end{enumerate}

\sidecaptionvpos{figure}{b}

\sidecaptionvpos{figure}{b}

%
%

\begin{algorithm}
	\caption{Interacted objects extraction}
	\begin{algorithmic}[1]
		\Procedure{Process}{video}
		\State Initialize \texttt{object\_dictionary} with boxes of interest and their features inf frame0.
		\State Initialize multiple trackers with those boxes.
		\For{frame in [frame1, frameN] in video}
		\State Extract boxes of interest in frame.
		\State Filter boxes of interest to reduce noisy frames. The result is \texttt{filtered\_boxes}
		\State Extract features for those \texttt{filtered\_boxes}.
		\State Retrieve label by doing match features of boxes of interest to features of objects in \texttt{object\_dictionary}. The result is retrieved boxes \texttt{d\_boxes} and retrieved labels \texttt{d\_labels}
		\State Track boxes by using trackers. The result is track boxes \texttt{t\_boxes} and their labels \texttt{t\_labels}.
		\State Pick boxes in \texttt{d\_boxes} that do not overlap with \texttt{t\_boxes} for creating new trackers. 
		\State Update \texttt{object\_dictionary} with \texttt{d\_boxes} and \texttt{t\_boxes}.
		\State Detect hands in the frame. The result is hand boxes \texttt{h\_boxes}.
		\State List object boxes in \texttt{d\_boxes}, \texttt{t\_boxes} that overlap with the hand boxes \texttt{h\_boxes}. The result is \texttt{o\_boxes}.
		\State \Return \texttt{o\_boxes}
		\EndFor
		
		\EndProcedure
	\end{algorithmic}
\end{algorithm}

Algorithm 1 shows the final part of preprocessing, which extracts the
list of objects used in each step.  The output of this algorithm is an
object association list that maps each working step to objects
involved in interactions in that step.  We bootstrap the algorithm by
using unsupervised object recognition on the frames, as described by
Pham et al.~\cite{pham2018unsupervised}  Only a subset of these objects may be used in any particular step.  The lists \texttt{o\_boxes}
and \texttt{h\_boxes} are two crucial elements for recognizing the
hand-object interaction used for object detection.  The \texttt{o\_boxes} are extracted from tracked boxes \texttt{t\_boxes}
and detected boxes \texttt{d\_boxes}. Tracked boxes are obtained by an
object tracker, while detected boxes come from objects with similar features in \texttt{object\_dictionary}, which is created and updated frame-by-frame by analyzing the workflow video.
The end results of PT is a temporal segmentation of the video into a sequence of 
steps, and a list of visual regions corresponding to objects associated with each step.  Note that the identities of 
the objects is not actually determined here.

\section{PTE\lc{ditor:} C\lc{orrecting} E\lc{rrors} \lc{in} P\lc{reprocessing}}
\label{sec:workfloweditor}
\begin{figure*}[htb]
	\centering
	\includegraphics[width=0.9\textwidth]{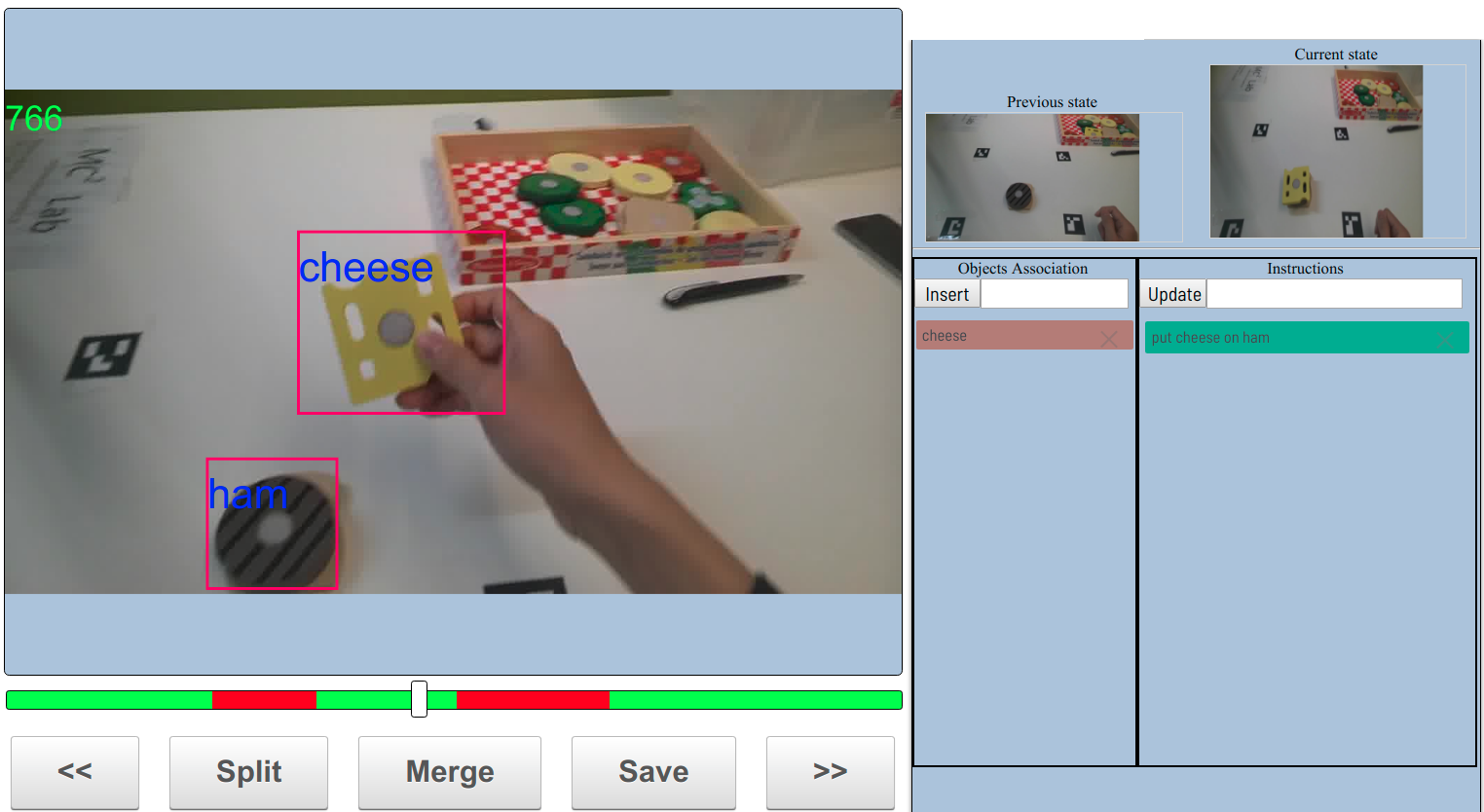}
	\caption{Graphical user interface of PTEditor}
	\label{fig:workflow_editor}
\end{figure*}

Because of the limitations of computer vision, the workflow extraction
process described in Section~\ref{sec:preprocessing} is imperfect.
For example, the object association list for a step may be missing an object, 
or include one in error.    Another possible error is wrong lull detection,
resulting in an incorrect demarcation of a working step.
Such errors  have to be
manually corrected by the development team before the extracted workflow can be used.
Figure~\ref{fig:workflow_editor} illustrates the GUI-based tool called
PTEditor that we have created for this purpose.  This tool provides
support for browsing and editing a workflow, as described below.

\textbf{Browsing:} A video can be loaded and viewed on the window in
the left panel.  The "\textless\textless" and
"\textgreater\textgreater" buttons at the bottom of that panel enable
quick navigation to the first frame of the previous, or next, working
step respectively.  Frame-by-frame advance through the video is also
possible.  A horizontal scroll bar presented under the window shows
segments representing working steps in different colors.

\textbf{Editing:} The "Split" and "Merge" buttons under the horizontal
scrollbar can be used to correct erroneous demarcations of a working
step.  As their names imply, these buttons can divide a step, or merge
two consecutive steps centered at the current frame.  For each frame,
the right panel shows the visual completion state and object
association list for the associated step.  Objects can be manually
added to or deleted from an association list.  Upon completion of
edits, the "SAVE" button at the bottom creates a JSON file that is the
external representation of the extracted workflow.

\begin{figure*}[t]
	\centering
	\includegraphics[width=0.95\textwidth]{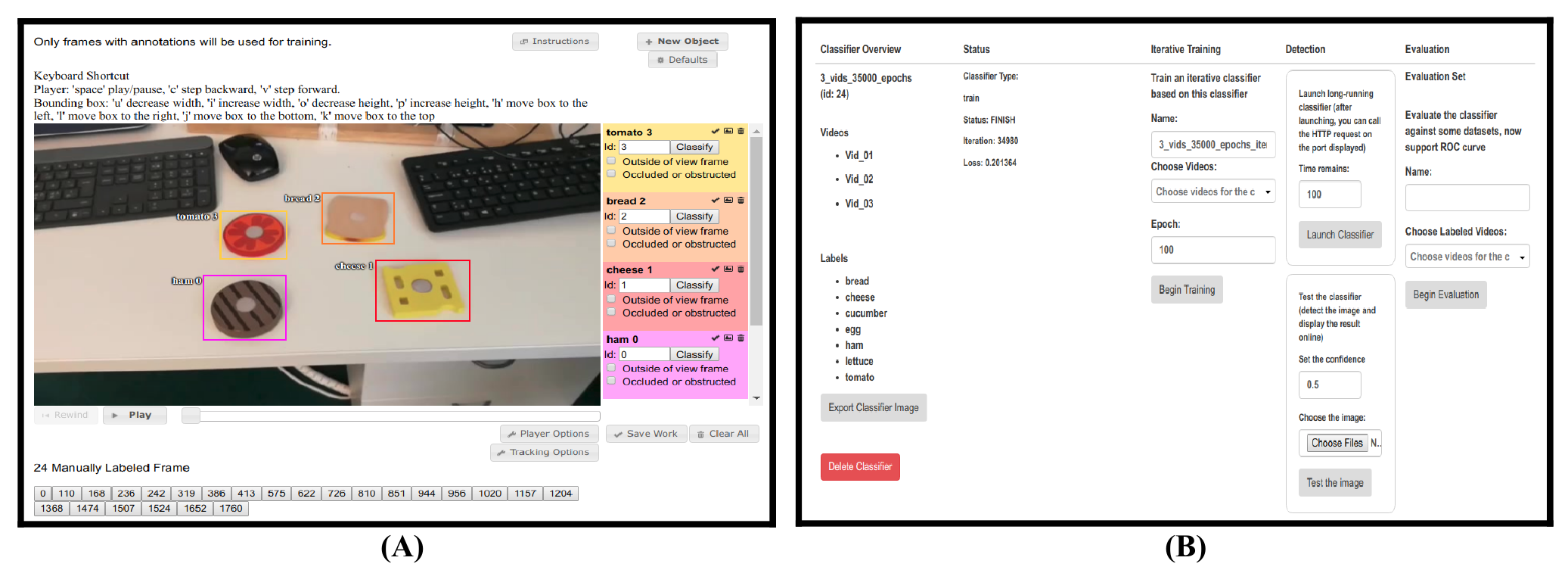}
	\caption{OpenTPOD GUI. (A) - Data labeling user interface. (B) - DNN training and testing user interface.}
	\label{fig:tpod_gui}
\end{figure*}

\section{O\lc{pen}TPOD: A\lc{n} O\lc{pen} T\lc{oolkit} F\lc{or}  P\lc{ainless} O\lc{bject} D\lc{etection}}
\label{sec:tpod}

The next component of the Ajalon toolchain is OpenTPOD (Open Toolkit
for Painless Object Detection).  The use of computer vision in a
Gabriel application relies on accurate detection of objects in the
association lists of working steps.  OpenTPOD enables the creation of an
accurate detector for these objects even if the development team lacks
machine learning and computer vision skills.  It does this by helping
to create a training set for a deep neural
network (DNN), and then training the DNN.  To reduce the size of the
training set required, OpenTPOD uses a technique called {\em transfer
  learning} that starts with a pretrained model for a set
of objects from a public dataset.  In addition to reducing labeling
effort for creating the training set, OpenTPOD also automates and
simplifies the process of training the DNN.

For each object in the union of object lists across working steps, the
developer first captures a few short videos of the desired object.
These are typically captured with a smartphone, from different viewing
angles and against multiple backgrounds.  The videos are uploaded via
a web browser into OpenTPOD.  A few minutes of video will contain
several thousand frames containing the object of interest.
Fortunately, as described below, only a handful of these frames need
to be manually labeled. This is done by drawing a bounding box around
the object in those frames.

Figure~\ref{fig:tpod_gui} A illustrates the labeling process.  First,
the developer finds the initial occurrence of the object in the video
and labels that frame.  Then, using a {\em correlation tracking
  algorithm},~\cite{danelljan2014accurate} OpenTPOD extrapolates the
position of the object (and hence, the bounding box) in subsequent
frames.  Because of imperfect tracking, the bounding box will drift
relative to the actual object in later frames.  When the developer
judges that the drift is excessive, he pauses OpenTPOD and explicitly
labels that later frame.  This reinitializes the position of the
bounding box for tracking in subsequent frames.  Then, OpenTPOD
continues its extrapolation until the drift is again judged to be
excessive by the developer.  This alternation of extrapolation and
explicit labeling is continued until the end of the video.  In our
experience, this approach of labeling followed by tracking can reduce
the number of frames that need to be manually labeled by a factor
of 10X~--~20X.

Next, OpenTPOD  performs a data cleaning and augmentation pass.  Because
of interframe correlations, many of the object examples may be close
to identical. OpenTPOD will eliminate the near duplicate examples, as
these will not help in the training process.  Optionally, data
augmentation can be employed. This creates more images by applying image manipulation techniques on the original image such as rotation, flipping, changes to the color scheme, and distortion. Such augmentation has been shown to help produce more robust object detectors.~\cite{shorten2019survey}

Finally, OpenTPOD can automate the transfer learning of a DNN model
using the collected dataset with the GUI shown in
Figure~\ref{fig:tpod_gui} B.  By default, OpenTPOD uses a
state-of-the-art Faster R-CNN VGG network~\cite{ren2015faster} pre-trained on the Pascal VOC
dataset,~\cite{Everingham15} though other network architectures can be
used as well. Negative examples are mined from the video background;
these are parts of the frames not included in the object bounding
boxes.  The training is started as a batch process that uses a
standard, scripted learning schedule, and generates both the final
Tensorflow model, as well as a Docker container with the executable
detector. 
Once training is complete, OpenTPOD provides download
links for the generated model and Docker container. 

Overall, OpenTPOD can greatly reduce both the labeling effort and
in-depth machine learning knowledge needed to effectively train and
deploy a DNN-based detector. We note that OpenTPOD is largely a
stand-alone tool that can be used separately from the rest of Ajalon.

\section{O\lc{pen}W\lc{orkflow:} B\lc{ringing} \lc{the} P\lc{ieces} T\lc{ogether}}
\label{sec:statemachine}
\begin{figure}[ht]
    \centering
    \includegraphics[width=0.8\textwidth]{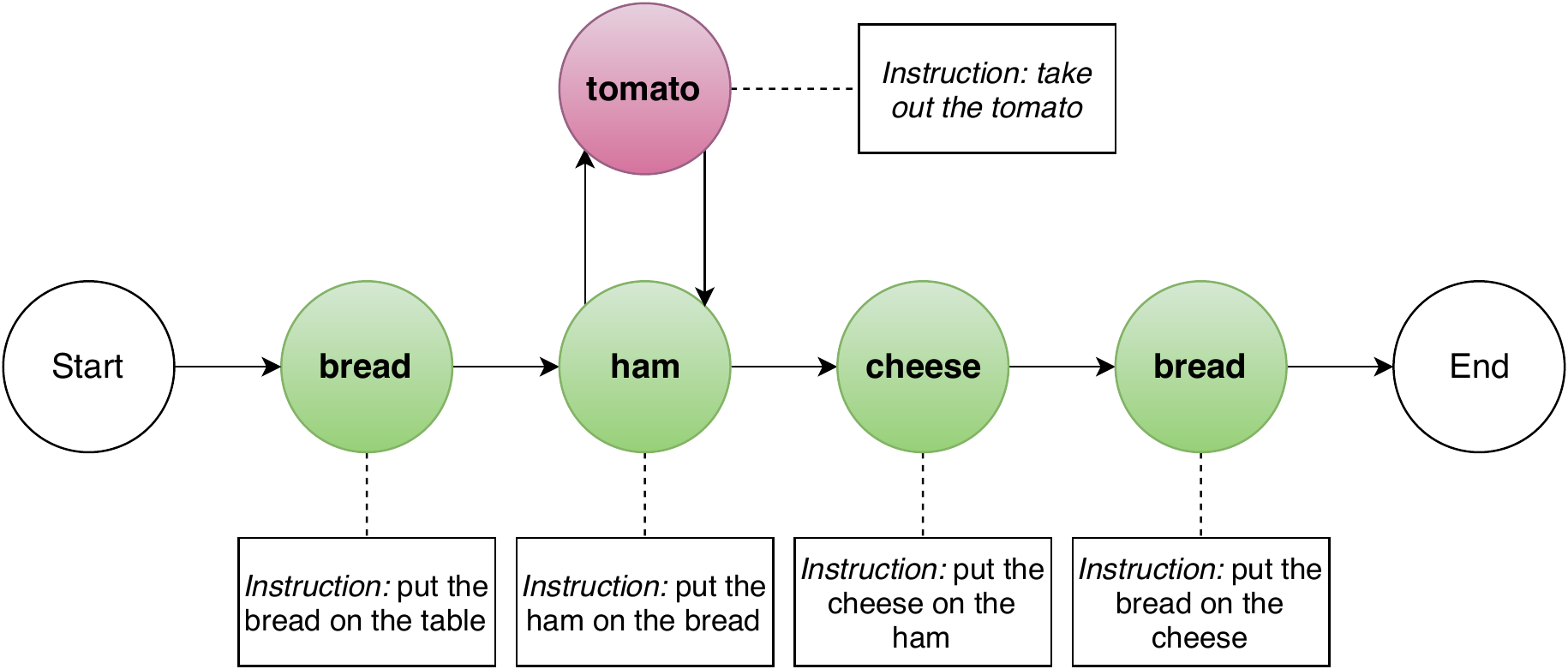}
    \caption{An example of finite state machine (FSM) for the application that assists in making a sandwich. White circles represent the start and end states of the process. Green circles represent correct states that help the user finish the task. The red circle represents an error state that requires the user to recover to the latest correct state to finish the task.}
    \label{fig:openworkflow}
\end{figure}

The final stage of Ajalon is OpenWorkflow, a finite state machine
(FSM) editor that automatically generates executable code for an FSM.
Each state can be viewed as a program counter value in the workflow of
the task.  At the start, the states of the FSM
correspond to the output of PTEditor.  The editing function of
OpenWorkflow can be used to add new states that, for example,
correspond to distinct error states resulting from frequently-committed
errors by users. The editing function can also be used to describe the
appearance of a state, using one or more object detectors that were
created by OpenTPOD.  As mentioned earlier, Ajalon assumes that the
state of progress of a task can be detected by its visual appearance.
For example, the user might put a piece of ham on top of a piece of
bread when making a sandwich.  The detection of a
``ham-on-top-of-bread'' object in a video frame indicates transition
into this state.  Figure~\ref{fig:openworkflow} shows a sample task
workflow as an FSM that can be edited using this tool.  The editing
process can also indicate the visual and verbal guidance provided to a
user as annotations on state transitions.  When the editing process is
complete, OpenWorkflow generates compiled code that implements the FSM
as an executable application.

\section{Evaluation}
\label{sec:evaluation}

We explore the effectiveness and usability of Ajalon through two sets
of questions.  There are four questions in the first set, each
pertaining to a different element of the Ajalon toolchain.  These are
answered via microbenchmarks in Sections~\ref{sec:pt_micro} through
\ref{sec:sme_micro}:
\begin{enumerate}
\item Is the automatic working-step segmentation of PT accurate?

\item Can a PT-extracted workflow be easily edited using PTEditor?

\item Can facility with OpenTPOD easily be acquired, and how good are its object detectors?

\item Is OpenWorkflow easy and effective to use for adding error
  cases, modeling workflow, and generating executable code?
\end{enumerate}
Questions in the second set pertain to the Ajalon toolchain as a whole,
and are answered in Section~\ref{sec:user_study} through a user study:
\begin{enumerate}
\setcounter{enumi}{4}
\item Do developers find Ajalon easy and effective to use?

\item Does Ajalon improve developer productivity?
\end{enumerate}

\subsection{A\lc{ccuracy} \lc{of} PT}
\label{sec:pt_micro}

\textbf{Approach:} To evaluate the accuracy of PT, we used the  Hape
assembly video dataset,~\cite{pham2018unsupervised}.  This dataset
consists of 40 10-minute videos of the process for building two toy
models.  The working steps are discrete actions that occur during
assembly.

\noindent\textbf{Metric:} To measure the effectiveness of automated working
step  segmentation, we used Boundary Detection Accuracy (BDA)
,~\cite{Wang:2005:AGP:1101149.1101309} which compares the automated
solution to a manual one using the following equation:

\begin{equation}
BDA = \frac{\tau_{db}\cap \tau_{mb}}{max(\tau_{db},\tau_{mb})} \in [0,1]
\end{equation}
where $\tau_{db}$ and $\tau_{mb}$ are automatically detected working
step boundary and the manually labeled one, respectively.

\noindent\textbf{Results.} The average BDA on the entire dataset was $0.85$,
confirming that the working-step segmentation algorithm effectively
detects steps in the video.  However, this result also indicates 
that significant manual effort is required to eliminate
the residual error.  We discuss potential improvements to PT in
the next section.
 
\subsection{Effectiveness of PTEditor}
\label{sec:we_micro}

\textbf{Approach:} A sandwich toy assembly video was used as an
editing benchmark.  One of the authors, who was very familiar with the
assembly process, played the role of a task expert.  He was given four
sample workflow videos, each showing a different variant of making a
sandwich.  These four variants consisted of the following items
between two slices of bread:
(a) cheese, on top of tomato, on top of ham, on top of cucumber;
(b) tomato, on top of ham, on top of lettuce, on top of cucumber;
(c) tomato, on top of cucumber, on top of egg; and, 
(d) ham, on top of egg, on top of cucumber. 
The task expert used PTEditor on two of the videos, and a separate 
documenting tool on the other two.

\noindent\textbf{Metric and Results.} Our metric is the amount of time needed
to complete workflow information for each working step. Without
PTEditor, it took approximately 4.5 minutes, whereas with PTEditor, it
took only 2.6 minutes, a savings of 42\%. With the support of
extracted workflow information from PT and PTEditor, the task expert
easily captured the process in the video and quickly documented
workflow information from it.

\subsection{Effort Saved by OpenTPOD}
\label{sec:tpod_micro}

\textbf{Approach:} Using two different tools and playing the role of a
developer, one of the authors built an object detector for 6 different
pieces of the sandwich toy mentioned earlier.  The two tools were
OpenTPOD and CVAT~\cite{opencvcvat} from OpenCV.  CVAT is a widely
used tool for data labeling, and is actually used for the labeling
function of OpenTPOD.  By holding the GUI for the labeling of a single
frame constant across the two tools, the comparison isolates the value
of OpenTPOD's mechanisms for optimizing labeling effort, and for
setting up the training.  As mentioned earlier, the mechanisms for
optimizing labeling effort include propagation of labels across frames
through tracking, and the elimination of training examples that are of
low value.

\noindent\textbf{Metrics:} Three metrics are relevant.  The first is labeled
objects per frame (\textit{lof}).  This is defined as the total number
of labels, divided by the number of frames in the videos.  The second
metric is the number of lines of code (\textit{loc}) that were written
by the developer to format annotations, train, and test the DNN. The
third metric is the total time in hours (\textit{hr}) that the
developer spent in writing these lines of code.

\noindent\textbf{Results:} 
Table~\ref{table:tpod_ben} illustrates the breakdown of metrics in 5 steps 
for the two methods.  The 5 steps include:
(a) collection of 5 videos that contain the 6 sandwich pieces of interest;
(b) labeling the videos using OpenTPOD
(c) formatting annotations for training
(d) training the DNN
(e) testing the DNN on a small test set of 5 images.
The results in Table~\ref{table:tpod_ben} show that OpenTPOD offers a
clear advantage.  It requires much less manual effort in labeling (fewer
than 40\% the number of labels), and does not require writing code.

\begin{table}[t]
	\centering
	\begin{threeparttable}
	\begin{tabular}{clcc}
		\hline
		& \textbf{Task / Method} & \textbf{without OpenTPOD} & \textbf{with OpenTPOD}  \\
		\hline
		(a) & Data collecting & \multicolumn{2}{c}{0.33 \textit{hr}}  \\
		\hline
		(b) & Data labeling & 4497~~/~~6356 \textit{lof} & 1695~~/~~6356 \textit{lof}   \\
		\hline
		(c) & Annotation formating & 104 \textit{loc}, 1.08 \textit{hr} & 0 \textit{loc}, 0 \textit{hr}  \\
		\hline
		(d) & Coding for training & 5 \textit{loc}, 0.58 \textit{hr} & 0 \textit{loc}, 0 \textit{hr} \\
		\hline
		(e) & Coding for testing & 7 \textit{loc}, 0.33 \textit{hr} & 0 \textit{loc}, 0 \textit{hr} \\
		\hline
	\end{tabular}
	\begin{tablenotes}
		\item \textit{lof}: manual labeled boxes over number of frames
		\item \textit{loc}: number of lines of code
		\item \textit{hr}: hour
	\end{tablenotes}
\end{threeparttable}
	\caption{Benefit of Using OpenTPOD.}
	\label{table:tpod_ben}
\end{table}

\subsection{Effort Saved by OpenWorkflow}
\label{sec:sme_micro}
\textbf{Approach:} To evaluate the benefit of using OpenWorkflow, we
conducted a case study.  One of the authors, an experienced
programmer, was provided with the workflow for the sandwich assembly
task and a pre-built object detector for all the relevant objects in
that task.  Using these inputs, his goal was to create a Gabriel
application for the task twice: once using OpenWorkflow, and once
without it.

\noindent\textbf{Metrics:} 
Three metrics are relevant. The first is the time spent in adding error states
to the workflow.  This is measured by total time in hours
divided by number of error states (\textit{hr/error}). The second metric is
the number of lines of code written to create the Gabriel
application (\textit{loc}).  The third metric is the time in hours (\textit{hr}) spent
in writing these lines of code, including the conceptual effort of
mapping the task workflow and error states to code.

\noindent\textbf{Results:} Table~\ref{table:smewm_ben} presents our
results, broken down into four components.  
Using OpenWorkflow consumes slightly more time to add error cases.
However, that modest increase is more than compensated by the fact
that zero effort is needed for the other three steps when using
OpenWorkflow.  The results thus show a clear savings when using OpenWorkflow.

\begin{table}[t]
	\centering
	\begin{threeparttable}
	\begin{tabular}{clccl}
		\hline
		& \textbf{Task / Method} & \textbf{non-OpenWorkflow} & \textbf{OpenWorkflow} \\
		\hline
		(a) & Adding Error cases & 0.21 \textit{hr/error} & 0.29 \textit{hr/error}\\
		\hline
		(b)  & Workflow modeling & 0.17 \textit{hr} & 0 \textit{hr}  \\
		\hline
		(c) & Programming & 352 \textit{loc}, 22.5 \textit{hr} & 0 \textit{loc}, 0 \textit{hr}  \\
		\hline
		(d) & Testing \& Bugs fixing & 9.5 \textit{hr} & 0 \textit{hr} \\
		\hline
	
	\end{tabular}
	\begin{tablenotes}
	\item \textit{hr/error}: hour per error 
	\item \textit{loc}: number of lines of code 
	\item \textit{hr}: hour
	\end{tablenotes}
	\end{threeparttable}
	\caption{Performance comparison between non-OpenWorkflow and OpenWorkflow method in modeling the workflow and programming the Gabriel application.}
	\label{table:smewm_ben}
\end{table}

\begin{figure}[t]
	\centering
	\includegraphics[width=0.8\textwidth]{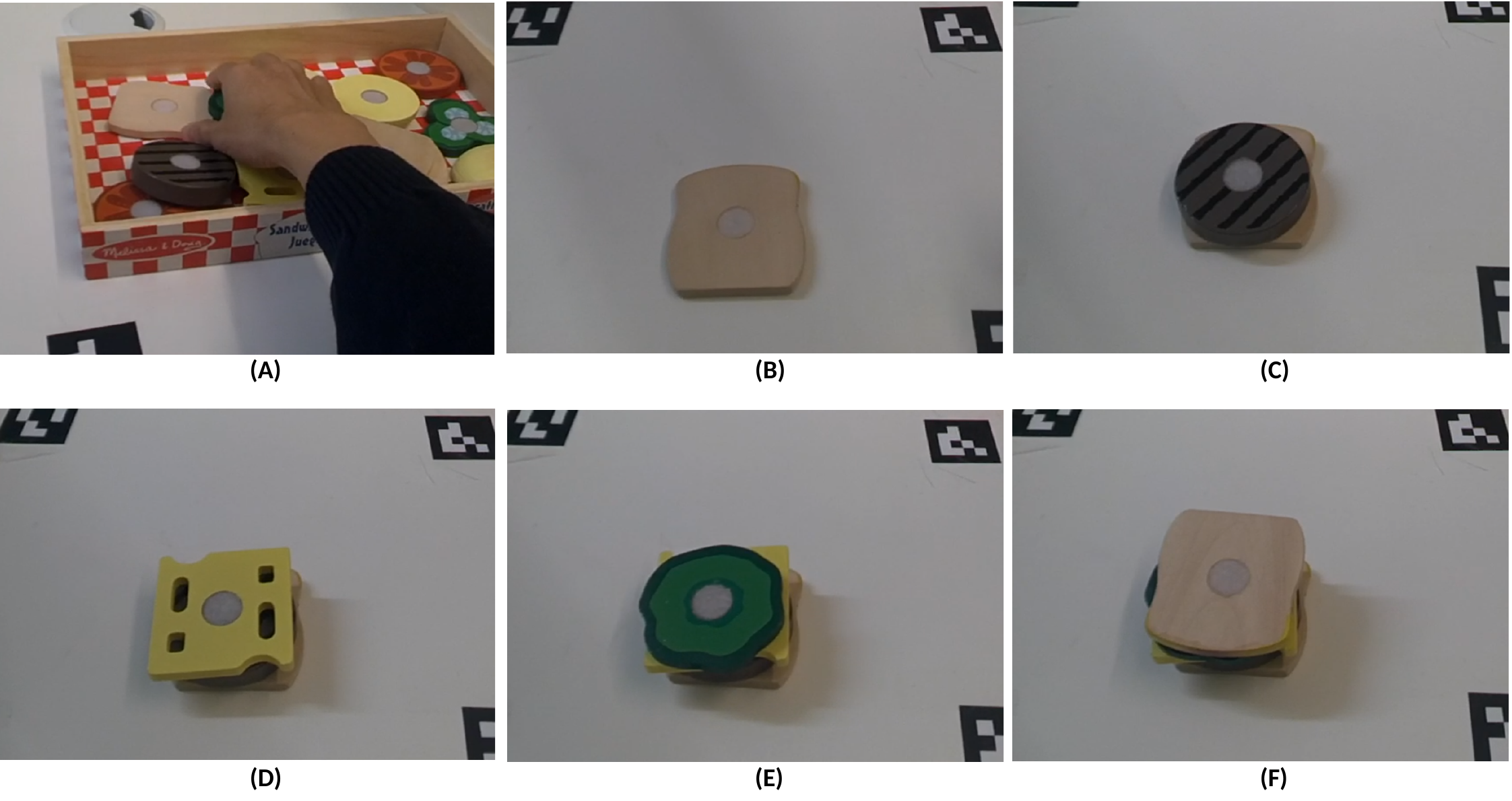}
	\caption{Workflow process of making a sandwich toy with the toy set. (A) Toy set box containing all components. (B-F) Assembly steps in order.}
	\label{fig:sandwich_process}
\end{figure}

\subsection{End-to-End User Study}
\label{sec:user_study}

To complement the microbenchmarks described in Sections
~\ref{sec:pt_micro} to \ref{sec:sme_micro}, we conducted a user study
of the full Ajalon toolchain.  This user study answers 
Questions 5 and 6 that were posed earlier.

\begin{SCfigure}
	\centering
	\includegraphics[width=0.65\textwidth]{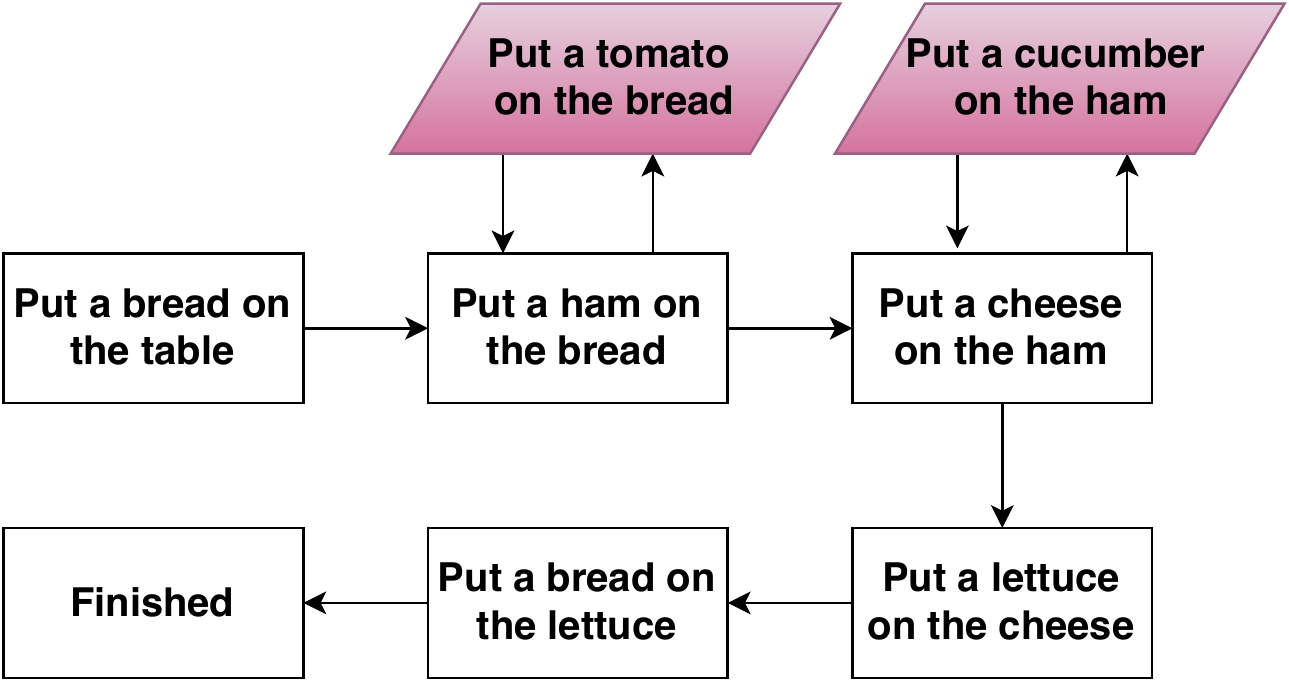}
	\caption{Test cases for sandwich application evaluation.  Normative steps for case 1 are in the white boxes; errors for case 2 and 3 are introduced as shown in pink.}  	\label{fig:AB_test_case}
\end{SCfigure}

\noindent\textbf{Approach:} 
Our user study asks subjects to write a WCA application with and without the Ajalon tools.
The target application should guide a user in making a sandwich 
from a toy kit, as described earlier.  We provide a
58-second first-person video of the \textit{making sandwich} process
to the subject.  Some key frames of this video are shown in
Figure~\ref{fig:sandwich_process}.  
The application produced by the subject should be capable of handling 
three test cases illustrated in
Figure~\ref{fig:AB_test_case}. In the first test case, the tester
performs the normative step-by-step process of sandwich building, as
presented in the original video.  This process corresponds to the
state machine represented by the white boxes in
Figure~\ref{fig:AB_test_case}. In the second and third test case, the
tester performed the normative process plus an error, shown as a red
parallelogram in the figure.  In the second test case, the error was
introduced after the first step: \textit{put a tomato on the bread}.
In the third test case, the error step occurred after the second
normative step: \textit{put a cucumber on the ham}. A functional
target application for making sandwiches should provide appropriate
instructions to users for all test cases.

\sidecaptionvpos{figure}{b}

\noindent\textbf{Subjects:}  
The subjects were eight students pursuing their Master's degrees in
Computer Science or Electrical Engineering at Aalto University in Finland. They were taking part
in a course that included three lectures on the concepts and
development process for producing a Gabriel application.  The average amount of
programming experience years for these subjects was $5.38$ years
($SD=2.13$).  Their fluency in different programming languages is
shown in Table~\ref{table:Programming_language}.  Their knowledge of
computer vision and deep learning is shown in
Table~\ref{table:background}, with values interpreted as follows:
(a) ``No experience:'' subject has no knowledge about the area;
(b) ``Beginner:'' subject has taken some courses related to the area;
(c) ``Intermediate:'' subject has studied or worked in the area for at least 3 years;
(d)``Advanced:'' subject has studied or worked in the area for at least 5 years.
We divided our subjects into two groups, junior and intermediate,
with 4 subjects in each.  Intermediate developers had at least 6
years of programming or were experienced with both computer vision and
machine learning, while junior ones had less than 6 years of
programming experience and 1 year of computer vision or machine
learning study.

\begin{table}[t]
	\centering
	\begin{tabular}{lllll}
		\hline
		Programming language & C/C++ & Python & Java & Others\\ \hline
		Proficient proportion & 63\% & 88\% & 38\% & 13\% \\
		\hline
	\end{tabular}
	\caption{Fluency in programming languages}
	\label{table:Programming_language}
\end{table}
\begin{table}
	\centering
	\begin{tabular}{lll}
		\hline
		Background Level& Computer Vision & Deep Learning\\ \hline
		No experience& 25\% & 37\% \\
		Beginner & 62\% & 50 \% \\
		Intermediate & 13\% &  13\% \\
		Advanced & 0 \% & 0\% \\
		\hline
	\end{tabular}
	\caption{Background in computer vision and deep learning}
	\label{table:background}
\end{table}

\noindent\textbf{Methods}. The subjects took part in two application
development conditions. The first required them to develop the
sandwich toy set application of Section 7.2 by using Ajalon. One week
later, the same subjects were asked to develop the same
application without using Ajalon. The time gap between the two
experiments is expected to reduce any memory bias on the same workflow
video. Moreover, any carryover learning from Ajalon should aid
performance in the manual version and reduce efficiency differences
between the methods. The total time for the user study was 3 weeks,
including one week for the Ajalon method, one week for the non-Ajalon
method, and one week for the time delay.

\begin{SCfigure}
	\centering
	\includegraphics[width=0.7\textwidth]{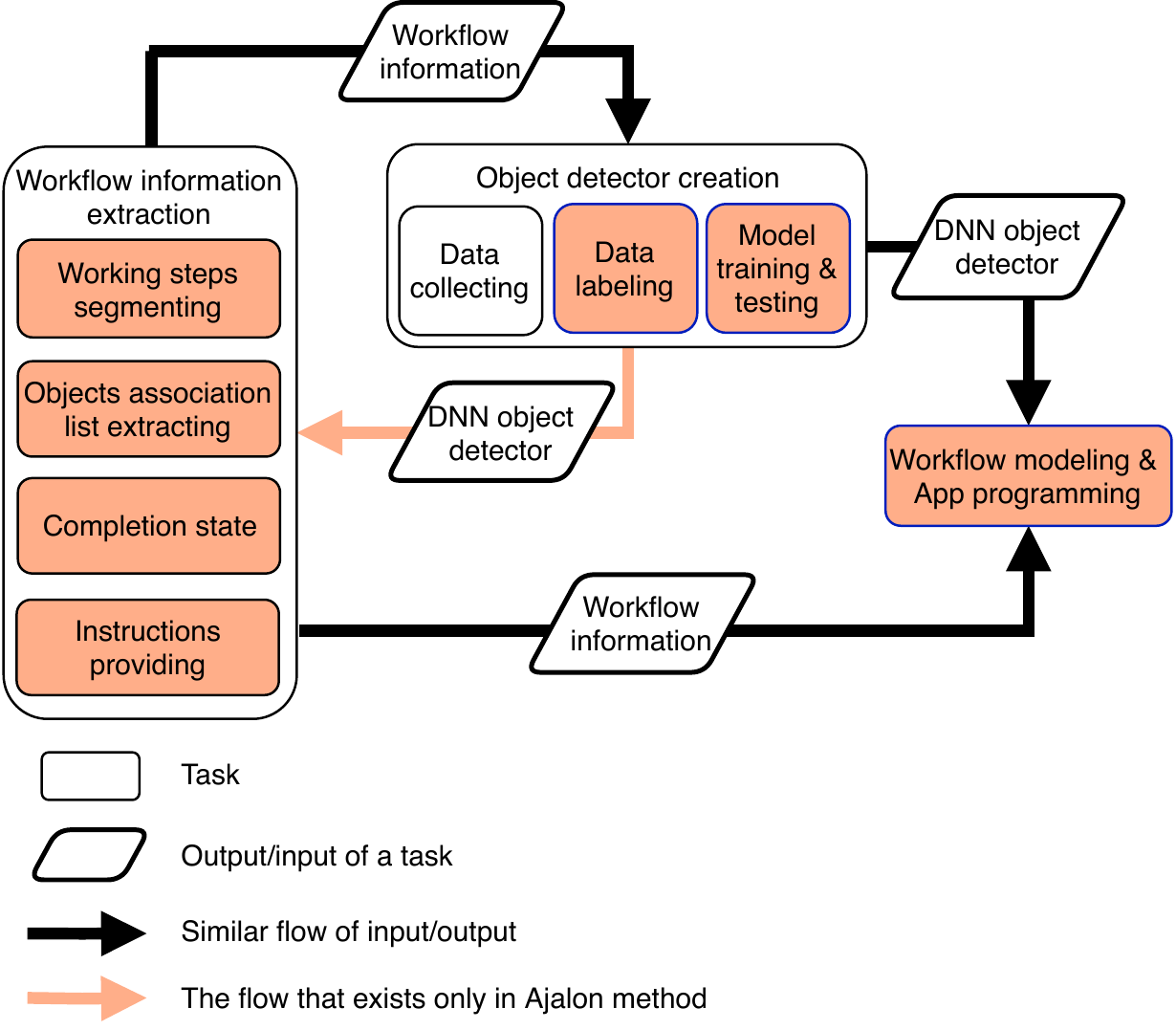}
	\caption{The Gabriel application development process with Ajalon and without Ajalon. Orange boxes represent tasks where the subject  used different tools with the Ajalon and non-Ajalon methods.}
	\label{fig:compare_developing}
\end{SCfigure}

As with the study described in Section 7.2, all subjects played
dual roles of task expert and developer. This is possible
because the task is simple: it can be mastered by a 5-minute training
session. Future work with more complex domains will test teams that
combine developers and task experts.  The subjects carried out
development as described in Section 7.2, except that for developing
the Gabriel application without Ajalon, we provided additional tools
for data labeling~\cite{russell2008labelme,opencvcvat} and DNN
training.~\cite{ren2015faster,redmon2018yolov3,lin2017focal}
Figure~\ref{fig:compare_developing} shows how the four major
development steps of the Gabriel development process differ in the
Ajalon and no-Ajalon cases.  Workflow extraction in the Ajalon case is
done using PT and PTEditor.  In the non-Ajalon case, this is done
using a video player.  Creation of object detectors in the Ajalon case
is done using OpenTPOD.  In the non-Ajalon case, this is done using a
manual setup of data labeling tools, training, and testing. The
implementation of the finite state machine for the Gabriel
application, including additions of error states, is done via
OpenWorkflow in the Ajalon case.  Since code is generated by
OpenWorkflow, no manual coding is necessary.  In the non-Ajalon case,
the representation of the finite state machine is created using a UML
drawing tool, and its translation into code is done manually.

\noindent{\bf Metrics:} As a basis for efficiency measures, subjects
were required to create working logs recording the time spent on their
specific steps of developing the application.  
A quantitative measure of savings is provided by the difference in time for the two conditions, relative to the with-Ajalon time baseline.  
\begin{equation} 
\label{eq:2}
\textit{savings factor} = \frac{\textit{Non-Ajalon time} - \textit{Ajalon time}}{\textit{Ajalon time}}
\end{equation}
Qualitative measures of the two conditions were obtained from a
post-experiment survey questionnaire that subjects filled out upon
completion of the study.  This questionnaire rated the usefulness,
ease, and enjoyment of each component of Ajalon on a discrete numeric
scale (1 = strongly disagree and 7 = strongly agree). PT, which was
fully automated, was not rated.  The survey questionnaire was
designed based on the guidance of Brooke et al.~\cite{brooke1996sus}
and Gordillo et al.~\cite{gordillo2017easy}

\sidecaptionvpos{figure}{b}

\begin{SCfigure}
	\centering
	\includegraphics[width=0.6\textwidth]{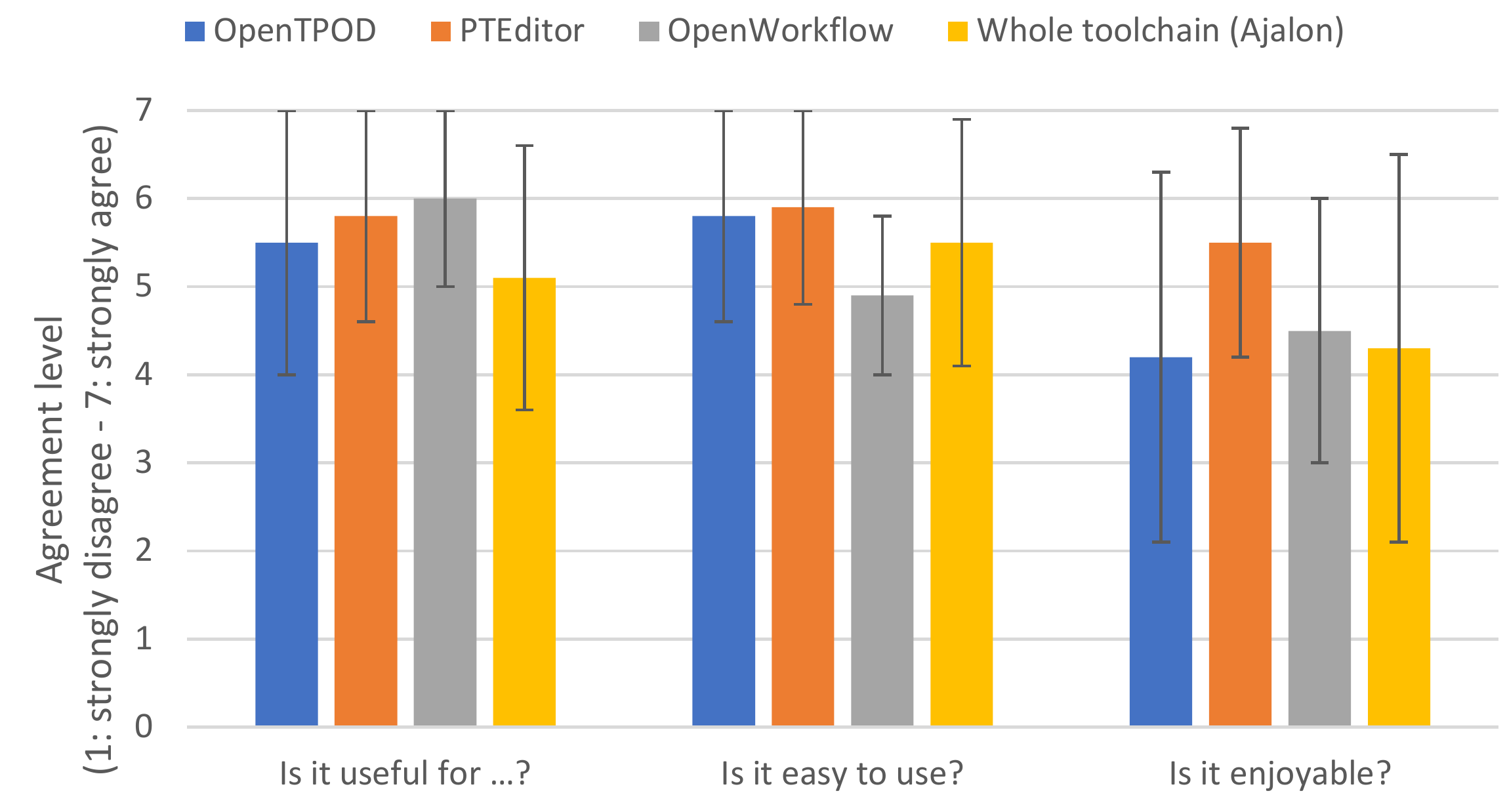}
	\caption{Rated user experience of using Ajalon usefulness, ease, and enjoyment; the scales ranged from 1 to 7.}
	\label{fig:usability_evaluation}
\end{SCfigure}

\noindent\textbf{Qualitative Results:}
Figure~\ref{fig:usability_evaluation} shows the results of the ratings
for each component of Ajalon.  Mean ratings ranged from 4 to 6 on the
7-point scale, indicating fairly high levels of usefulness (mean 5.1)
and ease of use (mean 5.5), and moderately high enjoyment (mean 4.2).
Ratings were similar across all components.  On the whole, these
results indicate that subjects perceived benefit from using
Ajalon in developing the application.

Some insight into the enjoyment scores being relatively low can be
gained from feedback given in the survey. Subjects found the
interface components, such as buttons and scroll bar, somewhat
annoying to use.  They also experienced some problems with the
tracking algorithm of OpenTPOD during labeling. These comments
underscore the importance of user experience design for Ajalon, as
enjoyment during use affects the performance in developing an
application.

\sidecaptionvpos{figure}{b}

\noindent\textbf{Quantitative Results:} The subjects' logs of time
spent in different stages of developing the application with and
without Ajalon revealed clear advantages for Ajalon condition. Means
by stage of application development for each of the two subject
groups, working with and without Ajalon, are shown in
Table~\ref{table:all_steps}.

\begin{table*}[t]
	\centering
	\begin{tabular}{llcccc}
		\hline
		\multicolumn{2}{l}{\textbf{MEAN TIME (SD) in hours}} & \multicolumn{2}{c}{\textbf{Junior}} & \multicolumn{2}{c}{\textbf{Intermediate}} \\
		\hline
		Task & Subtask & Without Ajalon & With Ajalon & Without Ajalon & With Ajalon \\
		\hline
		& Data labeling & 7.63 (1.6)& 3.25 (0.96) &  5.56 (0.52) &  3.13 (0.95)\\
		Object detector & Coding for training &  3.84(0.74) & 0 & 0.35 (0.10) & 0  \\
		 creation & Coding for testing & 0.88 (0.32) & 0 & 0.17 (0.07) &  0 \\
		\hline
		Workflow extraction & & 0.79 (0.17) & 0.35 (0.10) & 1.08 (0.24) & 0.42 (0.10) \\
		\hline
		Workflow modeling & & 0.75 (0.17)& 0.83 (0.53) & 0.90 (0.24)& 1.88 (1.11) \\
		\hline
		Other aspects & & 15.54 (0.95) & 0 & 7.98 (4.18) & 0 \\
		\hline
		All tasks & & 29.41(2.20) & 4.44 (1.21) & 15.91(4.78) & 5.42 (1.40) \\
		\hline
		
	\end{tabular}
	\caption{Time spent in component steps and the whole process of developing a Gabriel application for sandwich making, by junior and intermediate subjects when working with and without the aid of Ajalon. With Ajalon, OpenTPOD is used for creating the DNN-based object detectors, PT and PTEditor assist workflow extraction, and OpenWorkflow assists workflow modeling and code generating. Without Ajalon, the subjects can freely select their desired tools to achieve the required task.}
	\label{table:all_steps}
\end{table*}

\begin{SCfigure}
	\centering
	\includegraphics[width=0.65\textwidth]{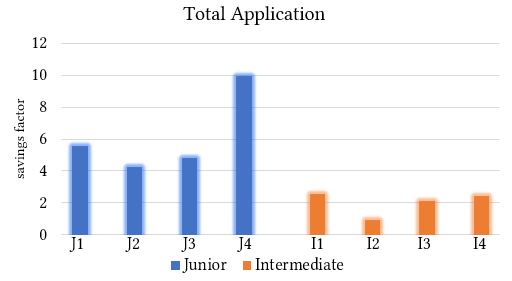}
	\caption{Time-saving proportion of using all tools of Ajalon for junior and intermediate subjects by \textit{savings factor} value based on \eqref{eq:2}. The higher the \textit{savings factor} is, the less time the subject spent on developing the application with the tool as compared to without it.}
	\label{fig:saving_proportion_all}
\end{SCfigure}

\begin{figure*}[ht]
	\includegraphics[width=0.95\textwidth]{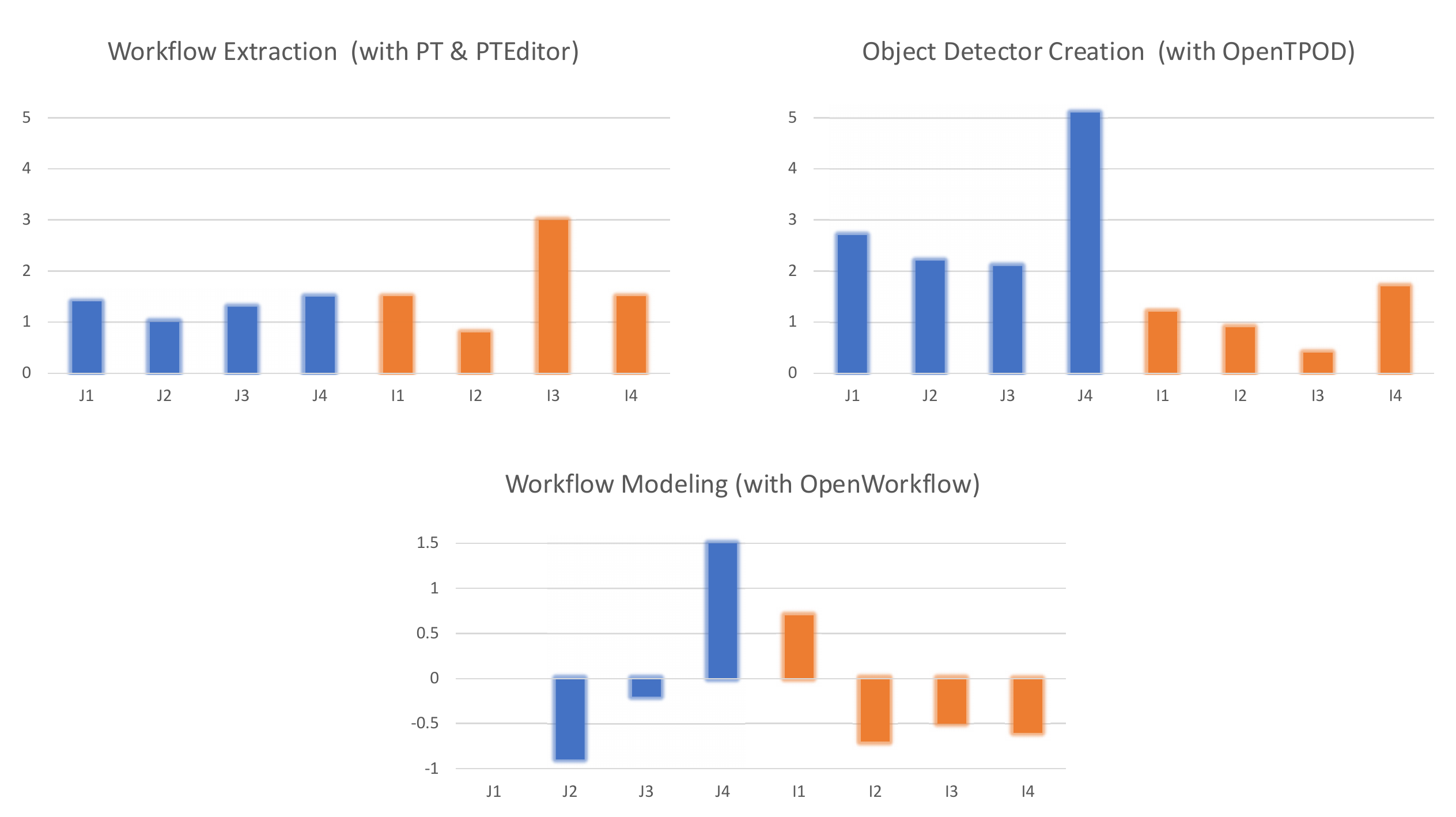}
	\caption{Time-saving proportion of using tools of Ajalon, or \textit{savings factor} value based on \eqref{eq:2}, for three steps for junior and intermediate subjects. With Ajalon, workflow extraction is supported by PT and PTEditor, object detector creation is supported by OpenTPOD, and workflow modeling is aided by OpenWorkflow.}
	\label{fig:saving_proportion}
\end{figure*}

Overall, subjects spent 22.7 hours in development without the aid of Ajalon, as compared to 4.9 hours with it.  
This \textit{savings factor} amounts to 3.6 overall and is greater for
junior-level subjects (5.6) than intermediates (1.9).  Another
implication of the table is that Ajalon aid leveled the playing field
for the two groups, bringing the average time for juniors and
intermediates to a few hours each.  Figure~\ref{fig:saving_proportion}
shows the savings proportion for individual subjects in each component
of application development and Figure~\ref{fig:saving_proportion_all}
for the application as a whole.

\sidecaptionvpos{figure}{b}  

\textit{Total application time:}  Although the statistical power is
limited by the small number of subjects, we conducted a mixed-model
ANOVA on the between-subject factor of expertise level (junior,
intermediate) and the within-subject factor, presence/absence of
Ajalon.\cite{gagnon2018sage} The ANOVA indicates whether each factor
has an effect, averaged over the other (the main effects), and whether
there are mutual dependencies (indicated by an interaction).  For the
overall application time, both main effects were significant, showing
the expected advantage for greater expertise, ($F_{1,6}=15.90$,
$p=0.007$), and for the presence of Ajalon,($F_{1,6}=221.51$,
$p<0.001$). In addition, there was a significant interaction term,
($F_{1,6}=36.92$, $p=0.001$), reflecting the finding that subjects
at the junior level of expertise had a greater overall advantage from
using Ajalon. This is evident from the individual-subject data in
Figure~\ref{fig:saving_proportion}, where all four junior subjects
show greater savings than any intermediate.

We next assess the contributions of the various components of Ajalon,
using the same ANOVA approach on the times for each component.  Only
statistically significant F-tests on the presence/absence of Ajalon
will be reported below.

\textit{Object detector creation from OpenTPOD:}  Essentially,
OpenTPOD halved the time for building the expected-object detectors.
As shown in Figure~\ref{fig:saving_proportion}, all subjects
showed a positive savings for this step, producing a significant
effect of Ajalon presence, ($F_{1,6}=69.75$, $p<0.001$). The magnitude
of the savings effect, 1.06 overall, was similar for the two
subject groups.  In comments, four of six subjects pointed to
the benefits of OpenTPOD's full pipeline DNN-based object detectors
creating process, which released them from minor time-consuming tasks
like data annotation conversion and complicated installation for
setting up the model training on the computer.

\textit{Workflow extraction using the PTEditor:} Note that workflow
extraction with Ajalon uses PT and PTEditor, but as the former is
fully automatic, the time essentially measures the use of PTEditor.
Although the average \textit{savings factor} for this step (1.43) is
equivalent to that of bulding object detectors with OpenTPOD, the
absolute time benefit of using Ajalon to extract workflow information
is small. On average, subjects spent 0.9 hour without PTEditor
while spending 0.4 hour with it, a significant difference,
($F_{1,6}=64.374$, $p<0.001$). Three of six subjects reported that
PTEditor helped them determine what information was needed.

\textit{Workflow modeling from OpenWorkflow:}  This component is used
to include error cases when describing and modeling workflow. In
contrast with other tools of Ajalon, OpenWorkflow produced no benefit
in time for this task. On average, subjects spent 0.8 hour for
describing the error cases and modeling the workflow without
OpenWorkflow, as compared to 1.4 hours with it. As illustrated in
Figure~\ref{fig:saving_proportion}, only 2 out of 6 subjects
showed a benefit from OpenWorkflow. Feedback from subjects
suggested two main reasons that lead to the results. The first is
that the interface of OpenWorkflow is not easy to use, particularly
because the visualized state machine window is not large enough to
manipulate the workflow for adding error states. The second reason is
that adding error cases is complicated because subjects must
report processing functions in each node and condition. Subjects
suggested that the tool should provide a \textit{copy\&paste} function
to reduce the text inputting effort. Although OpenWorkflow has low
points in enjoyment and ease, its output, which is the modeled
workflow, help the automatic code generator build a Gabriel
application within 1 second.

\textit{Other aspects:} This component refers to
programming that is required to complete developing a Gabriel
application with modeled workflow and qualified object detectors.
Because the OpenWorkflow of Ajalon appropriately models workflow,
OpenWorkflow can directly generate the target Gabriel application. In
contrast, programming without OpenWorkflow Ajalon requires several
hours, as reported in other implementation task of
Table~\ref{table:all_steps}. Specifically, Ajalon saved approximately
20.3 hours for junior subjects and 8.5 hours for intermediate
ones.  The reduction in manual effort from developers and experts in
building the required Gabriel application is a strong point for
answering Questions 5 and 6 above.

\section{L\lc{imitations} \lc{and} F\lc{uture} W\lc{ork}}

While Ajalon is clearly a helpful and effective toolchain, it is far
from perfect.  Our user study has given us valuable insights into
areas for improvement.  First, the GUIs of OpenTPOD and OpenWorkflow
can be improved.  In OpenTPOD, subjects wanted an estimate of the time
remaining until models would be fully trained. In OpenWorkflow, nearly
half the subjects expressed a desire to interact with the editor using
hotkeys and a command line. In addition to these improvements, we are
also considering the use of voice commands rather than mouse-and-keyboard interaction.

The workflow information format of Ajalon could be improved along the
lines suggested by Jelodar et al.~\cite{8563111}, who propose a more comprehensive
method that describes  both activities and object association lists. They found that
describing activities leads to a boost in the accuracy of modeling a cooking
activity. We plan to incorporate activity recognition into Ajalon, in
order to automate more of the process of generating WCA applications.

Our user study only had 8 subjects. An important reason for the
small number of subjects is the amount of time required for
participation by subjects.  As Ajalon improves and matures, we
anticipate the recruiting of users to be easier.  This will enable us
to run user studies that are larger, ideally closer to the 15-20
subjects that are typical for a user study of this
kind.~\cite{kim2019virtualcomponent, lo2019autofritz}

Our user study involved a task (building a sandwich) that almost all
subjects were familiar with.  This enabled them to play the dual roles
of task expert and developer.  In the future, we plan to run studies
of more complex and unfamiliar tasks, where the roles of a task expert and
a developer are distinct.

\section{C\lc{onclusion}}

WCA has gained considerable visibility in a short time, and is likely
to be transformative in education, health care, industrial
troubleshooting, manufacturing, and many other areas.  An important
enabler for widespread adoption of WCA will be the lowering of barriers
to creating applications of this genre.   Today, WCA development is
much more difficult and slower than, for example, web development.   Further, it requires
skills in areas such as machine learning and computer vision that are
not widespread among software developers.  Lowering this barrier is an
important step towards democratizing the creation of WCA applications.

To this end, we have created Ajalon, an authoring toolchain for WCA
applications that reduces the skill and effort needed at each step of
the development pipeline.  Starting from a video of an expert
performing a task, Ajalon automates the extraction of working steps.
Without writing a single line of code, it enables accurate DNN-based
object detectors to be created.  Once again, without having to write
code, it enables the generation of executable code for an FSM that models
the task.  Our evaluation shows that Ajalon is effective in
significantly reducing the barrier to creating new WCA applications.

\bibliography{ajalon2020}

\end{document}